\documentclass[sigconf]{acmart}
\AtBeginDocument{%
  \providecommand\BibTeX{{%
    \normalfont B\kern-0.5em{\scshape i\kern-0.25em b}\kern-0.8em\TeX}}}

\setcopyright{acmcopyright}
\copyrightyear{2018}
\acmYear{2018}
\acmDOI{XXXXXXX.XXXXXXX}

\acmConference[Conference acronym 'XX]{Make sure to enter the correct
  conference title from your rights confirmation emai}{June 03--05,
  2018}{Woodstock, NY}
\acmBooktitle{Woodstock '18: ACM Symposium on Neural Gaze Detection,
 June 03--05, 2018, Woodstock, NY} 
\acmPrice{15.00}
\acmISBN{978-1-4503-XXXX-X/18/06}

\usepackage{tablefootnote}
\usepackage{pifont}% http://ctan.org/pkg/pifont
\newcommand{\xmark}{\ding{55}}%

\definecolor{CJ-color}{RGB}{23, 225, 225}
\definecolor{VG-color}{RGB}{200, 30, 30}
\definecolor{JK-color}{RGB}{23, 200, 34}

\newcommand{\name}{EchoWrist}

\usepackage{color,soul}
\usepackage{booktabs} %much nicer tables
\usepackage{subfig} %for putting multiple subimages in one figure, with individual captions
\usepackage{graphicx}
\usepackage[utf8]{inputenc}
\usepackage[T1]{fontenc}

\usepackage{siunitx}
\newcommand{\hide}[1]{}

\sethlcolor{yellow}
\ifdefined\scifihidecomments
    \newcommand{\cheng}[1] {}
    \newcommand{\hyunc}[1] {} % Hyunchul
    \newcommand{\yax}[1] {} % Yaxuan
    \newcommand{\tuoc}[1] {} % Tuochao
    \newcommand{\sony}[1] {} % Songyun
    \newcommand{\ReviewerFeedback}[1] {}  
    \newcommand{\fran}[1] {}
    \newcommand{\rz}[1]{}
    \newcommand{\lw}[1] {}
    \newcommand{\mose}[1] {} % Mose
    \newcommand{\ke}[1] {} % Ke
\else
    \definecolor{burntorange}{rgb}{0.8, 0.33, 0.0}
    \definecolor{cadmiumgreen}{rgb}{0.0, 0.42, 0.24}
    \definecolor{cobalt}{rgb}{0.0, 0.28, 0.67}
    \definecolor{amber}{rgb}{1.0, 0.75, 0.0}
    \definecolor{fashionfuchsia}{rgb}{0.96, 0.0, 0.63}
    \definecolor{brightcerulean}{rgb}{0.11, 0.67, 0.84}
    \definecolor{frenchblue}{rgb}{0.0, 0.45, 0.73}
    \definecolor{darkslateblue}{rgb}{0.28, 0.24, 0.55}
    \definecolor{cerulean}{rgb}{0.0, 0.48, 0.65}
    \definecolor{darkpastelgreen}{rgb}{0.01, 0.75, 0.24}
    \newcommand{\hyunc}[1] { \textcolor{burntorange}{[{\hl{hyunc:}} {#1}}]}
    \newcommand{\yax}[1] { \textcolor{magenta}{[{\hl{yaxuan:}} {#1}]}}
    \newcommand{\tuoc}[1] { \textcolor{darkpastelgreen}{[{\hl{tuochao:}} {#1}]}}
    \newcommand{\sony}[1] { \textcolor{blue}{[{\hl{songyun:}} {#1}]}}
    \newcommand{\ReviewerFeedback}[1] { \textcolor{brightcerulean}{[{Reviewer Feedback:} {#1}}]}
    \newcommand{\fran}[1]{\textcolor{burntorange}{[{francois:}{#1}}]}
    \newcommand{\rz}[1]{\textcolor{teal}{[{Ruidong: }{#1}]}}
    \newcommand{\lw}[1]{\textcolor{fashionfuchsia}{[{liuwei:}{#1}}]}
    \newcommand{\mose}[1]{\textcolor{burntorange}{[{mose:}{#1}}]}
    \newcommand{\ke}[1] { \textcolor{red!55!yellow}{[{Ke:} {#1}}]}

\fi

\ifdefined\scifiblind
    \newcommand{\blind}[1]{[omitted for blind review]}
\else
    \newcommand{\blind}[1]{#1} %for camera ready (not blinded)
\fi

\begin{document}

\title[\name{}]{\name{}: Continuous Hand Pose Tracking and Hand-Object Interaction Recognition Using Low-Power Active Acoustic Sensing On a Wristband}

\author{Chi-Jung Lee}
\authornote{Both authors contributed equally to this research.}
\email{cl2358@cornell.edu}
\orcid{https://orcid.org/0000-0002-1887-4000}
\author{Ruidong Zhang}
\authornotemark[1]
\email{rz379@cornell.edu}
\orcid{https://orcid.org/0000-0001-8329-0522}
\affiliation{%
  \institution{Cornell University}
  \city{Ithaca}
  \state{New York}
  \country{USA}
}

% \affiliation{%
%   \institution{Cornell University}
%   \city{Ithaca}
%   \state{New York}

\author{Devansh Agarwal}
\email{da398@cornell.edu}
\orcid{https://orcid.org/0009-0005-1338-9275}
\author{Tianhong Catherine Yu}
\email{ty274@cornell.edu}
\orcid{https://orcid.org/0000-0002-3742-0178}
\affiliation{%
  \institution{Cornell University}
  \city{Ithaca}
  \state{New York}
  \country{USA}
}

% \affiliation{%
%   \institution{Cornell University}
%   \city{Ithaca}
%   \state{New York}

\author{Vipin Gunda}
\email{vg245@cornell.edu}
\orcid{https://orcid.org/0009-0000-5500-2183}
\author{Oliver Lopez}
\email{ojl23@cornell.edu}
\orcid{https://orcid.org/0009-0006-6328-193X}
\affiliation{%
  \institution{Cornell University}
  \city{Ithaca}
  \state{New York}
  \country{USA}
}

% \affiliation{%
%   \institution{Cornell University}
%   \city{Ithaca}
%   \state{New York}

\author{James Kim}
\email{jjk297@cornell.edu}
\orcid{https://orcid.org/0009-0002-3583-1114}
\author{Sicheng Yin}
\email{sy594@cornell.edu}
\orcid{https://orcid.org/0000-0002-0165-9750}
\affiliation{%
  \institution{Cornell University}
  \city{Ithaca}
  \state{New York}
  \country{USA}
}

% \affiliation{%
%   \institution{Cornell University}
%   \city{Ithaca}
%   \state{New York}

\author{Boao Dong}
\email{bd324@cornell.edu}
\orcid{https://orcid.org/0009-0001-1218-9336}
\author{Ke Li}
\email{kl975@cornell.edu}
\orcid{https://orcid.org/0000-0002-4208-7904}
\affiliation{%
  \institution{Cornell University}
  \city{Ithaca}
  \state{New York}
  \country{USA}
}

% \affiliation{%
%   \institution{Cornell University}
%   \city{Ithaca}
%   \state{New York}

\author{Mose Sakashita}
\email{ms3522@cornell.edu}
\orcid{https://orcid.org/0000-0003-4953-2027}
\author{Francois Guimbretiere}
\email{fvg3@cornell.edu}
\orcid{https://orcid.org/0000-0002-5510-6799}
\affiliation{%
  \institution{Cornell University}
  \city{Ithaca}
  \state{New York}
  \country{USA}
}

% \affiliation{%
%   \institution{Cornell University}
%   \city{Ithaca}
%   \state{New York}

\author{Cheng Zhang}
\email{chengzhang@cornell.edu}
\orcid{https://orcid.org/0000-0002-5079-5927}
\affiliation{%
  \institution{Cornell University}
  \city{Ithaca}
  \state{New York}
  \country{USA}
}

\settopmatter{authorsperrow=4}

\renewcommand{\shortauthors}{Lee and Zhang, et al.}

\begin{abstract}

Our hands serve as a fundamental means of interaction with the world around us. Therefore, understanding hand poses and interaction context is critical for human-computer interaction. We present EchoWrist, a low-power wristband that continuously estimates 3D hand pose and recognizes hand-object interactions using active acoustic sensing. EchoWrist is equipped with two speakers emitting inaudible sound waves toward the hand. These sound waves interact with the hand and its surroundings through reflections and diffractions, carrying rich information about the hand's shape and the objects it interacts with. The information captured by the two microphones goes through a deep learning inference system that recovers hand poses and identifies various everyday hand activities. Results from the two 12-participant user studies show that EchoWrist is effective and efficient at tracking 3D hand poses and recognizing hand-object interactions. Operating at 57.9mW, EchoWrist is able to continuously reconstruct 20 3D hand joints with MJEDE of 4.81mm and recognize 12 naturalistic hand-object interactions with 97.6\% accuracy.

\end{abstract}




\begin{teaserfigure}
    \centering
  \includegraphics[width=0.8\columnwidth]{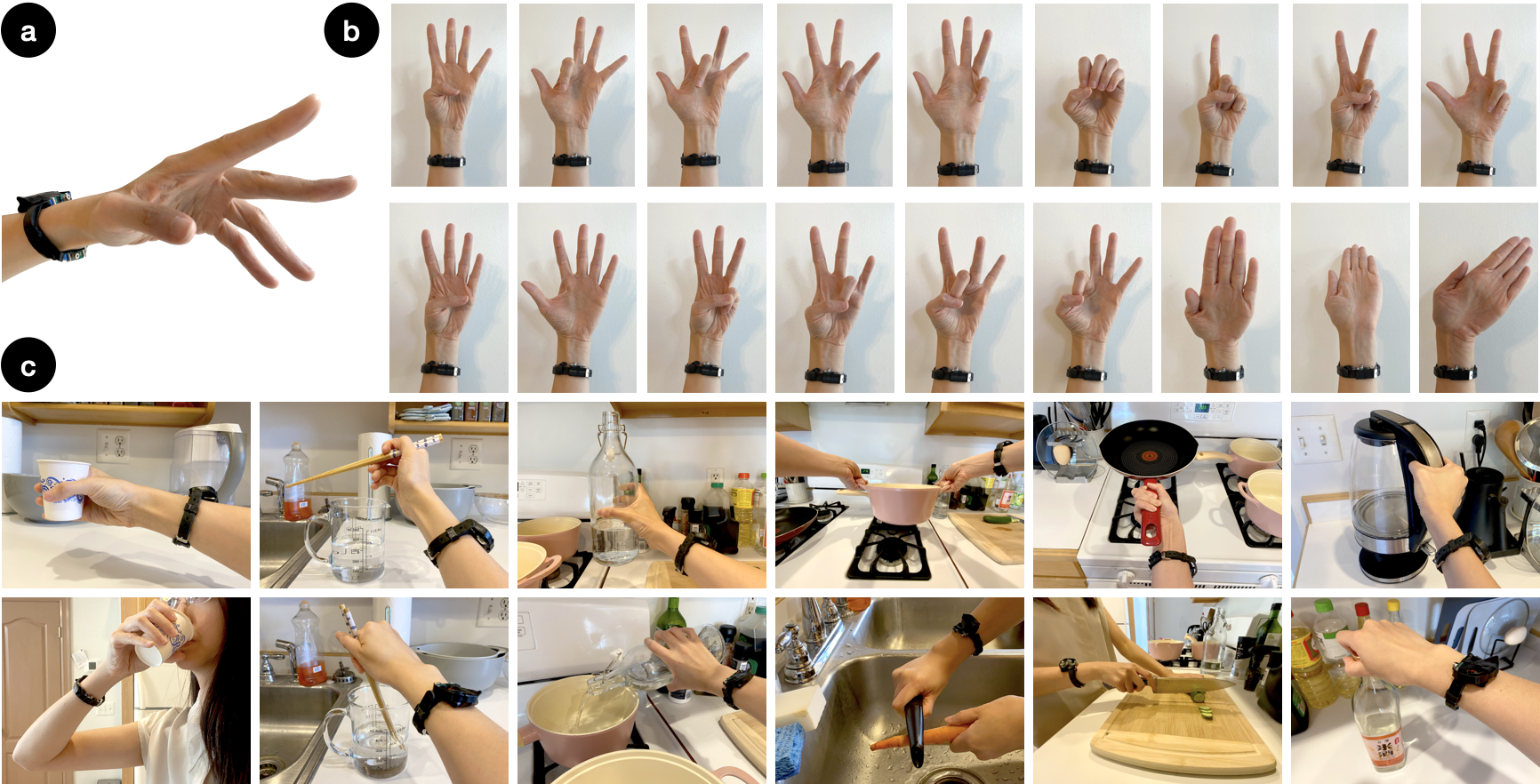}
  \caption{\name{} is a wristband that can understand 3D hand pose as well as hand object interactions. (a) The \name{} device. \name{} adopts a minimally-obtrusive design that keeps the device compact and low-profile. \name{} is able to (b) continuously track hand poses and (c) recognize various hand-object interactions.}
  \Description{(a) A picture of a hand with a black wristband on the wrist. The wristband has a similar shape of a commodity smart wristbands. (b) A grid of 30 figures demonstraing different hand gestures and hand-related activities, including: 1-5: bending one of the five fingers; 6-15: 10 gestures for 10 digits (0-9) in the American Sign Language; 16-18: the hand bends in three directions: flexion, extension and ulnar deviation; 19-24: a hand is seen holding a paper cup, a pair of chopsticks, a glass water bottle, a pot, a pan, and a kettle; 25-30: a hand is engaged with the following activities: drinking, stirring, peeling, twisting, chopping, and pouring.}
  \label{fig:teaser}
\end{teaserfigure}


\maketitle

\section{Introduction}
Human hands play an essential role in our daily lives. From nonverbal communication through gestures (e.g., Sign Language) to exploring our surroundings through touch and even grasping and manipulating objects, our hands serve as indispensable tools. Appreciating the significance of these activities involving our hands not only allows us to better understand our daily lives but also defines its practical applications, particularly in the context of human-computer interaction (HCI), which spans from passive context awareness to active input methods.

Building systems to track hand activities has been a long-standing challenge for the research community. This includes continuously capturing the 3D poses of hands and understanding the context of their interactions, e.g., what they are interacting with. The challenge arises due to 1) the highly flexible nature of hands and their numerous joints, 2) the potential occlusion of the fingers, and 3) the interaction with other objects that can change the shape of hands and obscure them from view. Consequently, understanding hand-object interaction poses a significant challenge since the system must possess knowledge of both the hand and the object. As a result, traditional hand-tracking solutions rely on external cameras to "see" the entire hand~\cite{oikonomidis2011full, leapmotion, kinect, realsense, lugaresi2019mediapipe, zimmermann2017learning, ge20193d, mueller2018ganerated, park2022handoccnet}. However, these methods are often intrusive, power-hungry, or require pre-setup, making them inconvenient for deployment in everyday life.

% Given the mobile nature of humans, it is essential to support the tracking of hand activities in mobile scenarios. 
In response to these challenges, researchers from the wearable community have proposed various solutions.
% , but striking a balance between device size, power consumption, comfortability, and performance remains a long-standing challenge.
%Many solutions deploy 
As skin-contacting sensors~\cite{liu2021neuropose, de2020real, raurale2018emg, sosin2018continuous, liu2021wr, zhang2015tomo, zhang2016advancing, kim2022ether, mcintosh2017echoflex, amento2002sound, deyle2007hambone, harrison2010skinput, fan2018your, laput2016viband} suffer from wearing discomfort, other solutions ~\cite{dementyev2014wristflex, lin2015backhand, rekimoto2001gesturewrist, truong2017low, truong2018capband, rudolph2022sensing, xu2015finger, chen2021vifin, xu2022enabling, iravantchi2019beamband} are limited to recognizing a pre-defined set of gestures without continuous pose tracking capability.
% which may not be comfortable to wear in everyday activities for long period of time as they require sensors to be tightly coupled to the skin. Most low-power and minimally-obtrusive solutions~\cite{dementyev2014wristflex, lin2015backhand, rekimoto2001gesturewrist, truong2017low, truong2018capband, rudolph2022sensing, xu2015finger, chen2021vifin, xu2022enabling, iravantchi2019beamband} can only recognize a set of pre-defined gestures without continuous tracking capability. 
In contrast, while wearable camera-based solutions~\cite{kim2012digits, yeo2019opisthenar, mcintosh2017sensir, wu2020back, hu2020fingertrak} support continuous tracking, they experience challenges in power consumption, e.g., 3.6 W\cite{devrio2022discoband}. This prevents the systems from full-day usage.  
Also, privacy concerns from bystanders are raised \cite{denning2014situ, o2023privacy, kudina2019ethics}.
% Recent wearable camera-based solutions~\cite{kim2012digits, yeo2019opisthenar, mcintosh2017sensir, wu2020back, hu2020fingertrak} have shown great promise in continuous tracking performance but come at the cost of higher power consumption~\cite{devrio2022discoband,hu2020fingertrak}, which operates at high power consumption level (e.g., 3.6W\cite{devrio2022discoband}). This makes them challenging to be immediately deployed on a wearable wristband with a very small battery capacity. 
% over 60 times higher than \name{}. 
Besides, most continuous solutions require an obtrusive device~\cite{kim2012digits, wu2020back, yeo2019opisthenar, maekawa2012wristsense, ohnishi2016recognizing, liu2021wr, liu2021neuropose} and typically focus on hand gestures without the ability to recognize hand object interaction. In summary, existing solutions exhibit at least one of the following limitations: 1) Discomfort from obtrusive form factor, 2) high power consumption, 3) lack of continuous tracking capabilities, 4) insufficient consideration of hand-object interactions, and 5) privacy concerns.
% Besides, most continuous solutions require a bulky device or sensors at a relatively high position~\cite{kim2012digits, wu2020back, yeo2019opisthenar, maekawa2012wristsense, ohnishi2016recognizing, liu2021wr, liu2021neuropose} and typically focus on hand gestures, has not demonstrated ability to recognize hand object interaction. 

To address these challenges, we introduce \name{}, a minimally obtrusive, low-power wristband designed to provide both continuous 3D hand shape tracking and a nuanced understanding of various hand-object interaction activities. \name{} utilizes active acoustic sensing, incorporating two pairs of compact speakers and microphones positioned in close proximity to the skin on each side of the wrist. The speakers emit inaudible frequency-modulated continuous waves (FMCW) directed toward the hand, and the resulting sound wave reflections and diffractions are captured by the wristband's microphones, creating distinct patterns corresponding to different hand poses. We then use a customized deep convolutional neural network (CNN) to continuously deduce the 3D hand poses represented by the 3D positions of 20 finger joints while also classifying various hand-object interactions. 

To evaluate \name{}'s performance in continuous hand pose tracking and hand-object interaction recognition, we conducted two user studies, each involving 12 participants. The results indicate that \name{} can continuously track 20 finger joints with a mean joint Euclidean distance error (MJEDE) of 4.81mm or mean joint angular error (MJAE) of 3.79\textdegree. Furthermore, \name{} exhibits impressive accuracy, achieving a recognition rate of 97.6\% across 12 diverse hand-object interactions, spanning static scenarios, such as firmly holding a cup, to dynamic actions involving movement, such as chopping. In addition, \name{} operates at a significantly low power consumption of just 57.9 mW,  with the sensing modules consuming only 10.0 mW, enabling full-day usage on standard smartwatches (e.g., 19 hours with 300mAh battery on an Apple Watch). Compared with previous work with continuous tracking capabilities~\cite{liu2021neuropose, liu2021wr, devrio2022discoband}, \name{} adopts a much smaller size and less obtrusive form factor.

This paper presents the following contributions:

\begin{itemize}
    \item We propose \name{}, a wireless, low-power, and low-profile wristband that can continuously track 3D hand poses and recognize hand-object interactions using active acoustic sensing.
    \item To our knowledge, \name{} is the first low-power and low-profile wristband that can both track 3D hand poses continuously and recognize hand-object interactions.
    % \item To our knowledge, \name{} achieves the most compact and minimally-obtrusive form factor in wristbands with continuous hand tracking capabilities.
    \item We evaluated \name{} with 3 user studies with 36 participants in total to demonstrate promising continuous hand tracking and hand-object interaction recognition capabilities.
    \item We presented the design considerations and iterations and further discussed the opportunities and challenges of deploying \name{} at scale.
\end{itemize}

\section{Related Work}
% \CJ{Hand tracking and recognizing hand activities have been the subject of extensive research over a long period of time\cite{sun2022human}.}
% In this section, we categorize research based on their \CJ{form factor and} sensing method, concluding with a discussion on acoustic sensing research, which is most pertinent to EchoWrist. Compared to similar wearable technologies in this space, EchoWrist demonstrates potential in its minimalistic form factor, reduced weight, and low power consumption.

% Compared with non-wearable methods, wearable solutions do not require users to be present in certain area (e.g., in front of cameras).
The wrist represents an advantageous location for hand sensing due to its inherent benefits, such as minimal interference with intricate finger dexterity and reduced susceptibility to external object occlusions.
We discuss previous wristbands on hand pose tracking, gesture recognition, and hand-object interaction. In addition, we provide a brief overview of sensing techniques using acoustic signals, which constitute our core sensing method.

\subsection{Hand Pose Tracking and Gesture Recognition}

Form factors other than wristbands exist in hand tracking/gesture recognition, such as gloves~\cite{perng1999acceleration} and rings~\cite{tapstrap2, zhou2022learning, tsai2016thumbring, parizi2019auraring, waghmare2023z}. However, wristbands usually create less interference with daily activities. With the increasing prevalence of smartwatches and smart bands, wristbands have more advantages for large-scale deployment.

\subsubsection{Discrete Gesture Recognition}
Discrete gesture recognition is relatively less challenging compared with continuous 3D hand tracking. Researchers have explored various sensors. A heavily-explored direction is using Electromyography (EMG), which captures electric signals from muscle movements \cite{de2020real, du2017semi, kerber2017user, quivira2018translating, raurale2018emg, sosin2018continuous}. Similar sensors include electric impedance sensing \cite{kim2022ether, zhang2015tomo, zhang2016advancing} and ultrasonic imaging \cite{mcintosh2017echoflex}. These technologies detect gestures through internal changes, but usually require skin contact and calibration. 
Another approach utilizes piezoelectric sensors \cite{amento2002sound, deyle2007hambone, harrison2010skinput} or high-frequency motion sensors \cite{laput2016viband} which capture bio-acoustic signals from wrist and finger movements. Technologies based on monitoring local shape changes use less invasive modalities such as pressure/flexors \cite{dementyev2014wristflex, lin2015backhand} and capacitive sensors \cite{rekimoto2001gesturewrist, rudolph2022sensing, truong2017demo, truong2017low, truong2018capband}, with some achieving ultra-low-power performance \cite{truong2018capband}.

Overall, skin-contacting technologies require sensors tightly attached to the skin, potentially causing discomfort over time. 
Therefore, contact-free hand-tracking technologies have garnered interest due to their comfort for extended use and promising performance potential. For instance, IMUs have been used on the palm/wrist~\cite{parizi2020auraring, tsai2016thumbring} or exclusively on smart devices~\cite{xu2022enabling} to recognize gestures. Methods exist such as proximity sensor arrays on the wrist \cite{gong2016wristwhirl, kim2007the} or thumb \cite{sun2021thumbtrak}, and vision sensors such as RGB \cite{wu2020back}, IR \cite{lim2023d, mcintosh2017sensir, yeo2019opisthenar}.
Still, due to their limited precision in capturing hand pose data, these methods usually recognize only a few gestures, which prevents a lot of potential applications.

\subsubsection{Continuous Hand Pose Tracking}
Recent advancement makes it possible to track hand poses continuously with a wristband.
Recent studies on EMG have shown promising results in recording continuous finger movements \cite{liu2021wr, liu2021neuropose}. However, EMG requires skin-contacting electrodes, which may not be comfortable for long-term wearing. In addition, many EMG-based methods usually place sensors at mid-forearm rather than a wristband, which may compromise comfort and convenience. Another promising direction is using wearable cameras, such as IR cameras~\cite{kim2012digits, yeo2019opisthenar}, thermal cameras~\cite{hu2020fingertrak} and depth cameras~\cite{devrio2022discoband}. However, they usually have significant power requirements and spacing constraints \cite{devrio2022discoband, yeo2019opisthenar, kim2012digits}, making them difficult to integrate into wearables such as smartwatches. For instance, DiscoBand~\cite{devrio2022discoband} operates at 3.6W while Digits~\cite{kim2012digits} requires a bulky camera on the palm's side.    Compared with previous work, \name{} operates at 57.9mW, and the highest point of the sensors is 5mm from the skin. \name{} provides a fully contact-free low-power solution that has a minimally obtrusive form factor with low-profile commercial speakers and microphones and provides continuous tracking capability.

\subsection{Hand-Object Interaction}
\subsubsection{Camera-Based Methods}
As cameras offer a wealth of data, including contextual information, the use of a wrist-mounted camera can simultaneously capture both hand posture and the surrounding environment. As a result, wrist-worn camera methods have been proposed to recognize daily activities \cite{maekawa2012wristsense, ohnishi2016recognizing}. However, to optimize data capture and minimize occlusion, the camera must be positioned at a certain distance from the skin, resulting in increased device thickness and potential discomfort. Additionally, there are concerns about the adequacy of privacy protection. To address these issues, DiscoBand \cite{devrio2022discoband} utilized multiple depth cameras while consuming high power and being bulky.

\subsubsection{Other Sensing Signals}
To minimize the occlusion issue, other signals were analyzed to understand hand activities. Fan et al. \cite{fan2018your} recognized in-hand objects via EMG. Rudolph et al. \cite{rudolph2022sensing} analyzed wrist topography to understand hand activities. However, these methods are sensitive to the grasping postures.

Since hands move to interact with the surroundings, motion-based methods were proposed. Using inertial sensors, \cite{zhang2022eatingtrak} detected the eating action. Using the multimodel method, Mollyn et al. \cite{mollyn2022samosa} and Bhattacharya et al. \cite{bhattacharya2022leveraging} sensed hand activities with lower power consumption. However, while the research demonstrated promising results, it was challenging to track static interactions that minimize movements.

On the other hand, some research has explored recognizing the activity based on the vibration profile from the contacting objects. Surface acoustic waves \cite{gong2020acustico} were used to sense the gestures against objects. ViBand \cite{laput2016viband} passively utilized accelerometers to capture bio-acoustic signals, while VibEye \cite{oh2019vibeye} actively propagated the vibration. In addition, Laput et al. \cite{laput2019sensing} leveraged commodity smartwatches to capture passive bio-acoustic signals. These methods achieve plausible results in recognizing the in-hand objects. However, passive vibration methods suffer from recognizing objects with minimal vibrations, while active vibration methods constrain the hand postures.

In summary, \name{} excels in achieving hand-object recognition through a low-power and low-profile wristband design while minimizing constraints related to interaction types, objects, and postures.

\subsection{Active Acoustic Sensing}
% \textcolor{teal}{As sound waves are omnipresent on Earth and capable of propagating through diverse media, researchers have explored numerous systems utilizing these characteristics. Acoustic signals passively collected from the environment have been employed to develop systems capable of recognizing everyday activities \cite{mollyn2022samosa, iravantchi2023sawsense}. To enhance control over the gathered information, active acoustic sensing has been proposed. By emitting sound waves into the environment and analyzing the receiving signals, rich information can be collected.}

Active acoustic sensing emits sound waves using speakers and receives the reflected acoustic waves using microphones. These received reflected acoustic signals contain rich information, e.g., position, and shape, about the object that reflected the signals. This sensing principle has been widely used as "Sonar" in the past. Acoustic sensors are widely available on modern computing devices including smartphones, wearables, and smart speakers. Therefore, much prior research has explored using active acoustic sensing on these form factors to recognize human activities. Some researchers explored using active acoustic sensing through surface propagation, which can be used to recognize body contact on surfaces \cite{ono2013touch} or touch gestures if combined with air-borne signals \cite{sun2018vskin, gong2020acustico}. Recently, researchers have shown using active acoustic sensing to track hand gestures using the microphone and speaker arrays on a smart speaker \cite{li2022room}, track facial expressions \cite{li2022eario} on earphones, understand silent speech \cite{echospeech} on glasses, recognize the which finger is interacting with a smartwatch \cite{kim2022sonarid}, or enabling new attachment on phones for new interaction \cite{laput2015acoustruments}.

The work aligned most with our work are the ones that use active acoustic sensing to sense hand gestures. FingerIO \cite{nadakumar2016fingerio} tracks 2D fine-grained finger movements around a smartphone or smartwatch. WristAcoustic \cite{huh2023wristacoustic} achieves gesture recognition for authentication on smartwatches. AudioGest \cite{ruan2016audiogest} enables gesture recognition on a laptop, tablet, or smartwatch. BeamBand \cite{iravantchi2019beamband} achieves gesture recognition on a smartwatch both within-session and across-session. While these systems exhibit promising results in gesture recognition, Ring-a-Pose \cite{yu2024ring} takes a step further by achieving continuous hand pose tracking using reflected information from the palm and fingers on a ring. 

In contrast, \name{} is the first wristband to use active acoustic sensing to continuously track hand poses and recognize hand-object interaction. It demonstrates a unique sensing principle that uses acoustic sensing to capture the shape and position of the wrist contour and surrounding objects, which can be learned by ML to infer hand poses and recognize hand-object interactions.

\section{Sensing principles and Design Considerations}

\name{} employs active acoustic sensing as its primary sensing method. In this approach, we use speakers to emit inaudible sound waves and microphones to receive the reflections of these emitted waves. These sound waves propagate from the wrist toward the palm, fingers, and the surrounding environment. The skin and surfaces of nearby objects act as the reflection medium for these sound waves. The waves undergo reflection and diffraction, eventually reaching the microphones.
Distinct hand shapes and the varying characteristics of the surrounding environment lead to different signal paths, resulting in complex multipath echo patterns, which could be distinguished with customized echo profile analysis and deep learning pipelines. 
% \name{} uses two pairs of speakers and microphones mounted on the top and bottom sides of the wrist, respectively. 

To facilitate this process, the final prototype of \name{} features two pairs of speakers and microphones strategically positioned on the top and bottom sides of the wrist, respectively. The inclusion of these two pairs of sensors allows for the comprehensive capture of echoes from both sides, thereby providing a wealth of information about hand gestures and interactions.

In order to determine the optimal design for \name{} in efficiently tracking 3D hand poses and recognizing hand-object interactions, a series of pilot studies were conducted. We aimed to explore the limits of the sensing principle and identify the optimal setup that balances device obtrusiveness, power consumption, and performance. Overall, two design considerations were proposed:

\textbf{(1) Minimally-obtrusive:} Given that wearable devices are typically worn throughout the day and come into prolonged contact with the user, ensuring comfort is a prioritized consideration in the design. Additionally, as our aim is to introduce a wristband, it's essential to prevent the device from interfering with daily hand-related activities. It is also vital to consider the social acceptability of the wearable device and how it can be blended with current wearables.
In summary, to optimize comfort and enhance the potential for integration with commercial smartwatches and wristbands, maintaining a minimally-obtrusive design that sits close to the skin is important. This includes using small and non-skin-contacting sensors and keeping the system small and low-profile.

% \textbf{(1) Low-Profile:} Given that wearable devices are typically worn throughout the day and come into prolonged contact with the user, ensuring comfort is a prioritized consideration in the design. Additionally, as our aim is to introduce a wristband, it's essential to prevent the device from interfering with daily hand-related activities. It is also vital to consider the social acceptability of the wearable device and how it can be blended with current wearables. In summary, to optimize comfort and enhance the potential for integration with commercial smartwatches and wristbands, maintaining a low-profile design that sits close to the skin is important.

\textbf{(2) Low-Power:} Wearable devices are generally worn for extended periods. Moreover, in the case of continuous tracking, they need to remain operational throughout. Therefore, it is crucial to consider power consumption during design.

% \textbf{(3) Compact:} To avoid disruption to everyday hand activities and improve overall comfort while wearing the device, it is advisable to maintain the close proximity of rigid electronic components. Additionally, for the sake of ease of maintenance, it is always beneficial to have these electronics positioned closely together. Overall, a compact design of the sensor placement is recommended.

Our goal was to find the balance between these design considerations and the sensing performance. To achieve the goal, the sensor type, number, and layout were examined in the studies.

\subsection{Speaker Type and Position}

\begin{figure*}[h]
  \centering
  \includegraphics[width=2\columnwidth]{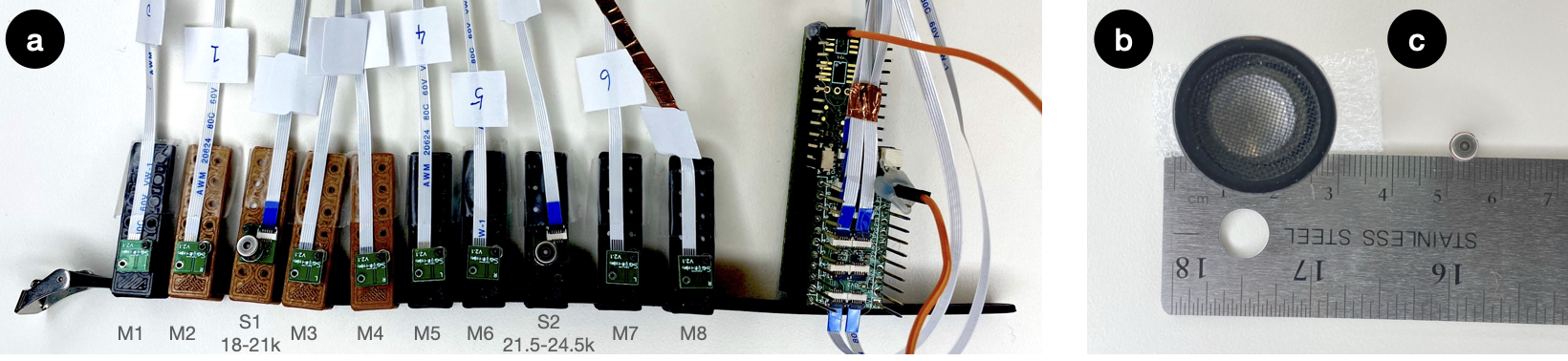}
  \caption{Pilot studies were conducted with (a) an experimental prototype. (b) An ultrasonic transducer was used for the initial design, while (c) a commodity speaker was chosen for the later prototype.}
  \Description{(a) A photo of an experimental prototype. 10 pieces of 3D-printed racks with multiple mounting holes are connected to a black silicone band. 8 small microphone boards and 2 small speaker boards are installed on the racks, respectively. A Teensy 4.1 module is by its side. The sensor boards and the Teensy module are connected with while FPC cables. (b) A picture of a large ultrasonic transducer and a small speaker againset a metal ruler. The diameter of the ultrasonic transducer is about 30mm while the small speaker is about 5mm.}
  \label{fig:design-hardware}
  \vspace *{-1.2em}
\end{figure*}

An experimental prototype based on Teensy 4.1~\footnote{\url{https://www.pjrc.com/store/teensy41.html}} with 3D printed sensor mounts was built to explore sensor configurations (Fig.~\ref{fig:design-hardware} (a)). The prototype allowed us to manipulate the sensor layout easily. We connected 2 speakers and 8 microphones to the prototype.

\begin{figure*}[h]
  \centering
  \includegraphics[width=2\columnwidth]{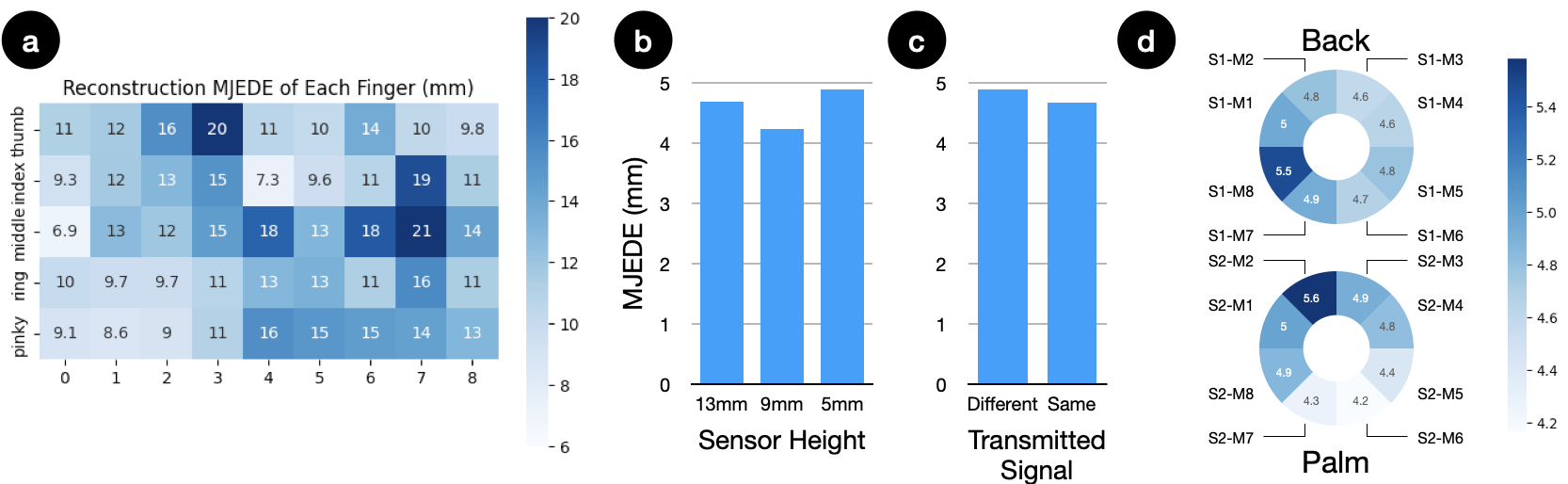}
  \caption{We explored the performance of (a) each finger in reconstruction mean joint Euclidean distance error (MJEDE) when the speaker was placed at different positions. Note that the numbers on the x-axis represent positions evenly distributed around the wrist. The center of the palm side is represented as 0, the one to its right as 1, and so forth. System MJEDE while adjusting (b) the height of the sensors, (c) the transmitted signal on the speakers, and (d) the combination of speaker and microphone were also examined. The number and position of the speakers and microphones can be referred to Figure \ref{fig:design-hardware} (a).}
  \Description{(a) A colored grid map with color saturation in correspondence to values. Title: Reconstruction Error of Each Finger. Vertical axis top to bottom: thumb, index, middle, ring, pinky. Horizontal axis (sensor position) left to right: 0-8. Values: row thumb: 11, 12, 16, 20, 11, 10, 14, 10, 9.8; row index: 9.3, 12, 13, 15, 7.3, 9.6, 11, 19, 11; row middle: 6.9, 13, 12, 15, 18, 13, 18, 21, 14; row ring: 10, 9.7, 9.7, 11, 13, 13, 11, 16, 11; row pinky: 9.1, 8.6, 9, 11, 16, 15, 15, 14, 13. (b) Bar chart. Title: Sensor Height. Vertical axis: MJEDE (mm). Series and values: 13mm - 4.7, 9mm - 4.2, 5mm - 4.9. (c) Bar chart. Title: Transmitted Signal. Series and values Different Signals - 4.9, Same Signal - 4.7. (d) Colored grid map. Title: S \& M. H axis: S1, S2, V axis: M1, M2, M3, M4, M5, M6, M7, M8. Values: column S1: 5, 4.8, 4.6, 4.6, 4.8, 4.7, 4.9, 5.5; column S2: 5, 5.6, 4.9, 4.8, 4.4, 4.2, 4.3, 4.9.}
  \label{fig:design-results}
  \vspace *{-1.2em}
\end{figure*}

We started with ultrasonic transducers as the signal emitter (Fig.~\ref{fig:design-hardware} (b)) for their excellent ultrasonic acoustic characteristics. We first tried to identify the optimal position for the speaker. We performed a grid search on the speaker position while doing single-finger movements (bending one of the five fingers at a time). One of the researchers collected the data and trained the model as described in Section \ref{sec:deep-learning}, and the reconstruction error was calculated according to the method outlined in Section \ref{sec:evaluation-metric}. Results (Fig.~\ref{fig:design-results} (a)) indicate that speaker positions under the palm yield the best performance. To maximize the information we can get with an additional speaker, we placed a second speaker in its opposite position above the back of the hand to obtain more diverse information from both sides of the hand.

In these experiments, we also realized that these transducers have strong directionality, which means that the emitted sounds are mostly directed to the fingers directly facing the transducer. These fingers had excellent performance, while other fingers did not work well. This inspired us to choose omnidirectional speakers so that sound waves can travel in all directions, allowing us to place speakers in less-obtrusive positions while still covering most fingers. We ended up using a commodity speaker, which comes in a smaller size as well (Fig.~\ref{fig:design-hardware} (c)).

\subsection{Sensor Height}
Next, we experimented with different sensor heights. We conducted a quantitative study on the sensor height. We chose three heights, placing the sensors 13mm, 9mm, and 5mm from the skin (measured from the skin surface to the outermost point on the sensor). With the last configuration, the lower edge of the speaker was almost touching the skin (the diameter of the speaker is 5mm). Two researchers conducted the experiments, doing a series of pre-defined single-finger movements and complex finger gestures, respectively. The average performance of the three speaker heights was very close (Fig.~\ref{fig:design-results} (b)). Since no strong decrease in performance was observed, we chose 5 mm sensor heights for the user study. This allows \name{} to adopt a low-profile form factor that can be easily integrated into commercial smartwatches or wristbands.

\subsection{Number of Sensors}
We then moved to optimize the number of sensors. We separated the emitted signals from the two speakers with different frequency ranges (18-21kHz and 21.5-24.5kHz). In this way, we could easily separate paths coming from the two speakers on each microphone. The MJEDE was calculated using data collected from a single pair of a speaker and microphone. The layout of 2 speakers and 8 microphones is specified in Fig.~\ref{fig:design-hardware} (a), where S1 was placed on the back of the hand, and S2 was under the palm. Two researchers experimented with using different channels and their combinations. When only using one of the 16 channels, results (Fig.~\ref{fig:design-results} (d)) indicate that microphones close to the speakers yield slightly better results, but other positions also work decently. We attribute this to the use of omnidirectional speakers combined with the use of echo profiles to maximally preserve information. Specifically, having the back speakers improves performance when the hand was bent outwards. We placed speakers at both the back and the palm sides of the hand to maintain reliable performance when the wrist is at different angles. To maximize the benefit of using 2 speakers, we decided to use 2 microphones that are close to the speaker. Compared with having more sensors, the combination of 2 speakers and 2 microphones can be easily achieved as digital audio interfaces such as Inter-IC Sound (I$^2$S) usually come with stereo audio.

We notice that the cross-path signals (signal traveling from the palm speaker to the back mic, and vice versa) were ignorable compared with direct-path signals. Another experiment on using the same frequency on the two speakers versus using different frequencies (Fig.~\ref{fig:design-results} (c)) confirmed that no significant performance drop could be observed. Allowing the two speakers to send the same signal can save half of the bandwidth, which can save hardware cost and size by using mono-channel audio amplifiers instead of stereo ones. Therefore, we decided to emit the same signal on the two speakers in our system.
\section{Implementation}
\subsection{Hardware Implementation}

\begin{figure*}[ht]
  \centering
  \includegraphics[width=2\columnwidth]{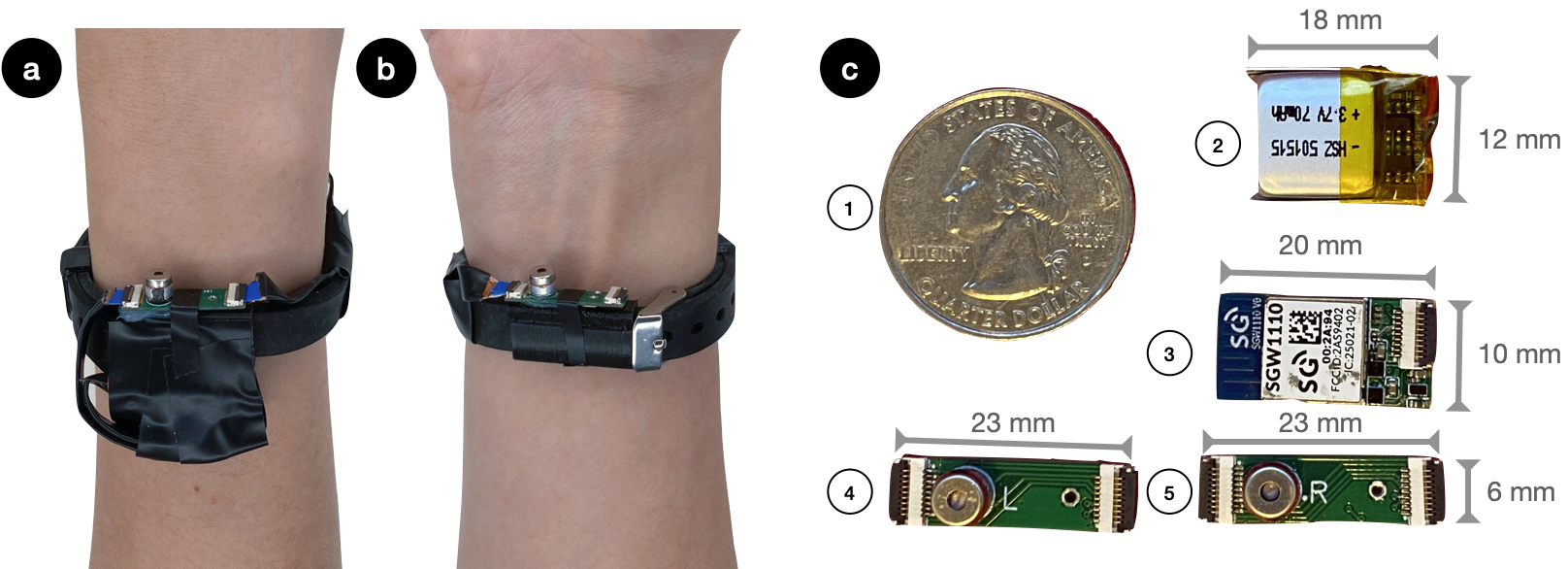}
  \caption{Hardware of \name{}. (a) (b) Wearing \name{} at the wrist. All components are mounted on a silicone wristband. (c) Customized PCBs for the microcontroller module and the sensing module. (1): US Quater Dollar coin. (2): 3.7V 70mAh LiPo battery. (3): Customized PCB with SGW1110 module. (4) \& (5): Sensor module with speaker and microphone.}
  \Description{(a-b) A hand is wearing a black wristband. Viewed from the back's and palm's sides. The sensor modules with speaker and microphone are visible. (c) Photos of different modules along with dimension markings. (1) US Quater Dollar coin (2) A 70mAh rectangle LiPo battery, size 18x12mm. (3) Customized PCB with SGW1110 module, size 20x10mm. (4-5) Customized PCB with sensors, marked with L and R, size 23x6mm.}
  \label{fig:hardware}
  \vspace *{-1.2em}
\end{figure*}

\name{} uses two pairs of speakers (OWR-05049T-38D) and microphones (ICS-43434) mounted on a low-profile silicone band (Fig.~\ref{fig:hardware} (a), (b)). We designed customized PCBs for the sensing module (Fig.~\ref{fig:hardware} (c)). The two sensing modules are connected via a flexible printed circuits (FPC) cable and then connected to the customized microcontroller module. The microcontroller board includes an SGW1110 module (with nRF52840 microcontroller), two MAX98357A audio amplifiers (in the study, only one was used), plus a power management module with a TPS62743 buck regulator. The entire system is powered by a LiPo battery. The sensing module and the microcontroller module are both attached to 3D-printed cases that can slide along the silicone band to fit different hand sizes. The cases were printed with thermoplastic polyurethane (TPU) so that they were soft and comfortable. The weight of the sensing module and the microcontroller module is 0.7g and 1.2g, respectively. The weight of the entire system, including the battery, is 16.8g. While collecting data, the microcontroller drives the speakers to emit sounds and collects echoes from the microphones. The collected data can be saved on the microSD card with an extended socket. To support real-time data collection, the collected data can also be transmitted to a smartphone (Xiaomi Redmi Note 10 Pro) via Bluetooth low-energy (BLE) operating at 800kbps. The captured data are truncated to 8 bits to save bandwidth in this case. No performance degradation was observed between using full 16 bits and truncated 8 bits.

\subsection{Power Signature}
Designed for compact wearable devices such as smartwatches and wristbands, \name{} aims to provide a low-power solution. We examine the power signature of \name{} using a CurrentRanger~\footnote{\url{https://lowpowerlab.com/guide/currentranger/}}. Results show that when the system is on with BLE transmitting data at 800 kbps, the power consumption is 57.9 mW (3.86V, 15.0mA).

At 57.9mW operation power, \name{} can easily last a full day with a common battery size of smartwatches (e.g., Apple Watch Series 8 has around 300mAh battery size~\footnote{\url{https://www.xda-developers.com/apple-watch-series-8-and-ultra-battery-size/}}, which should last 19 hours). If \name{} is integrated into the existing hardware of the smartwatch/wristband, the microcontroller's base power consumption could be saved since the sensors only operate at 10 mW, leading to an even longer battery life. Note that this calculation did not consider the power consumption of the operating system on a smartwatch.

\subsection{3D Hand Pose Ground Truth Acquisition}

\begin{figure*}[h]
  \centering
  \includegraphics[width=2\columnwidth]{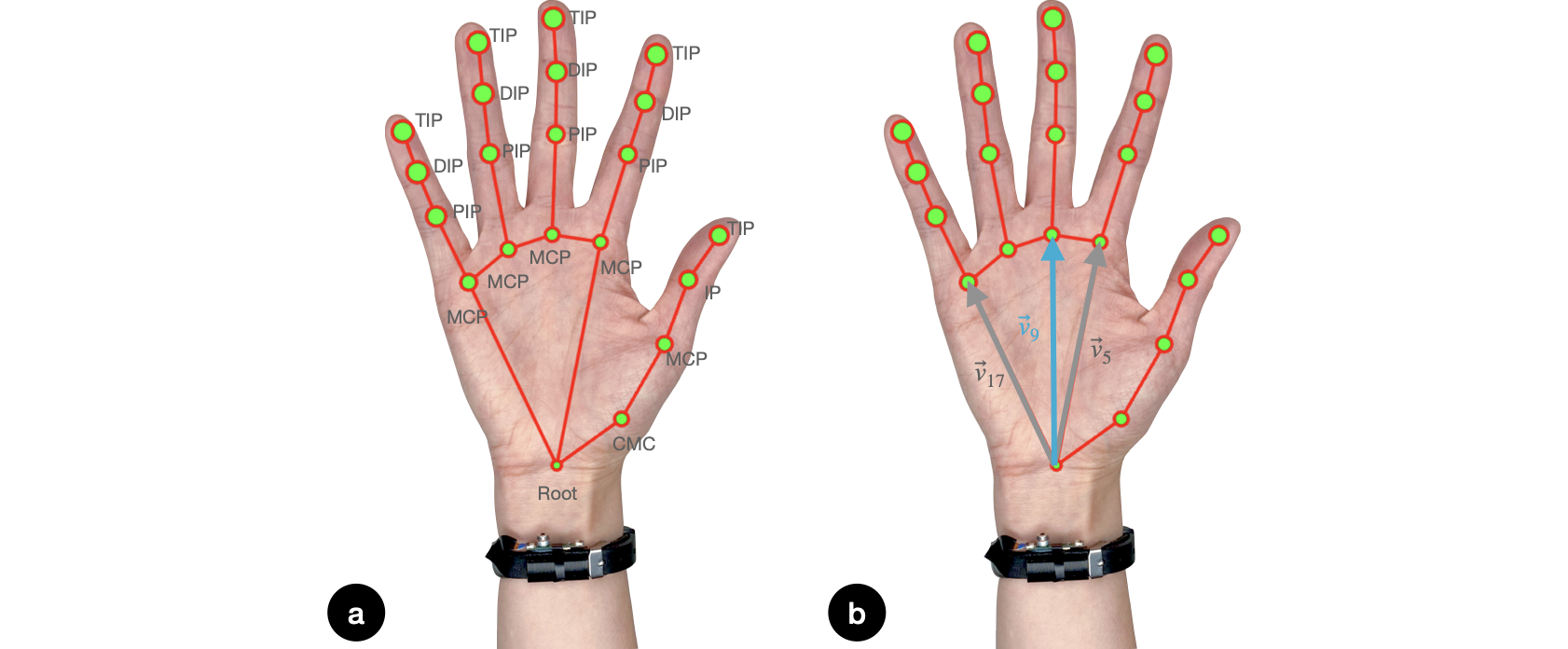}
  \caption{3D hand pose ground truth annotation. (a) 21 joints detected by MediaPipe. (b) Important vectors used during ground truth normalization. $\vec{v_9}$ is the wrist-to-palm vector used to represent the orientation of the hand. The plan defined by $\vec{v_5}$ and $\vec{v_{17}}$ is used to align the detected hand with the reference hand. The actual length of $\vec{v_{17}}$ is measured in the image against the size of the sensing module to uniform the hand size in each frame.}
  \Description{(a) A photo of an open hand with black wristband on the wrist. All joint names (Root, CMC, MCP, IP, PIP, DIP, TIP) are marked. (b) Same as (a) except no text markings and 3 vectors are marked: from root to MCP of index: $\vec{v_5}$, from root to MCP of middle: $\vec{v_9}$, from root to MCP of pinky: $\vec{v_{17}}$.}
  \label{fig:mediapipe}
  \vspace *{-1.2em}
\end{figure*}

We used MediaPipe~\cite{lugaresi2019mediapipe} to acquire the ground truth of 3D hand shape, which has been widely used in prior projects ~\cite{kim2022ether, devrio2022discoband}.
% \rz{maybe skip the part of leapmotion}
% To obtain the ground truth for 3D hand shape, we need a system that does not interfere with the acoustic sensing system. Computer vision-based methods are suitable for this purpose due to their reliable performance and simple setup. We compared Leap Motion \cite{leapmotion} as used in FingerTrak~\cite{hu2020fingertrak}, WR-Hand~\cite{liu2021wr} and MediaPipe \cite{lugaresi2019mediapipe} as used in~\cite{kim2022etherpose}. We noticed that Leap Motion could not provide as reliable tracking as MediaPipe did, especially when there was an occlusion between fingers. This is also one of the major reasons that FingerTrak and WR-Hand did not evaluate continuous tracking on complex gestures with occlusion. Furthermore, using Leap Motion requires the arm to be parallel to the table, which is more tiring over a long time. Therefore, we chose MediaPipe as the ground truth acquisition system. 
With MediaPipe, the shape of the hand is represented by 21 joints, including the wrist (Fig.~\ref{fig:mediapipe}(a)). Each joint is represented by 3D coordinates. While recovering the hand shape, we predict the coordinates of the 20 joints, excluding the wrist, which is set as the origin. While recovering the wrist rotation, we predict the wrist-to-palm vector as illustrated in Fig.~\ref{fig:mediapipe}(b).

\subsubsection{Ground Truth Normalization}

For the data collection of 3D hand pose tracking, we asked users to remount the device between sessions and allowed them to relax and move around. This causes variances in the relative position and orientation between the hand and the camera. In addition, MediaPipe may not capture the size of the hands reliably due to a lack of objects to compare. To fix these issues, we developed an algorithm to normalize the ground truth. For each frame, we first used MediaPipe to extract the positions of all hand joints. We then calculated the surface of the palm represented by vectors $(\vec{v_5}, \vec{v_{17}})$, where $\vec{v_5}$ is the vector pointing from the wrist to the metacarpophalangeal joint of the index finger (joint 5 in Figure~\ref{fig:mediapipe}(b)) and $\vec{v_{17}}$ is the vector pointing from the wrist to the metacarpophalangeal joint of the little finger (joint 17 in Figure~\ref{fig:mediapipe}(b)). We calculated the rotation matrix that rotates this plane to the reference plane $(\hat{\vec{v_5}}, \hat{\vec{v_{17}}})$ extracted from the reference posture (Figure~\ref{fig:mediapipe}(b)). We chose this plane because it remains largely static even when the user is performing complex gestures. For each participant, we also measured the length of $\vec{v_{17}}$ as against the size of the sensing module. We normalize each frame so that the length of $\vec{v_{17}}$ is equal to the measured value, as illustrated in Figure~\ref{fig:mediapipe}(b). While recovering the wrist rotation, ground truth normalization was not applied so that the wrist rotations were faithfully recorded.

\subsection{Data processing and deep learning pipelines}

\begin{figure*}[h]
  \centering
  \includegraphics[width=2\columnwidth]{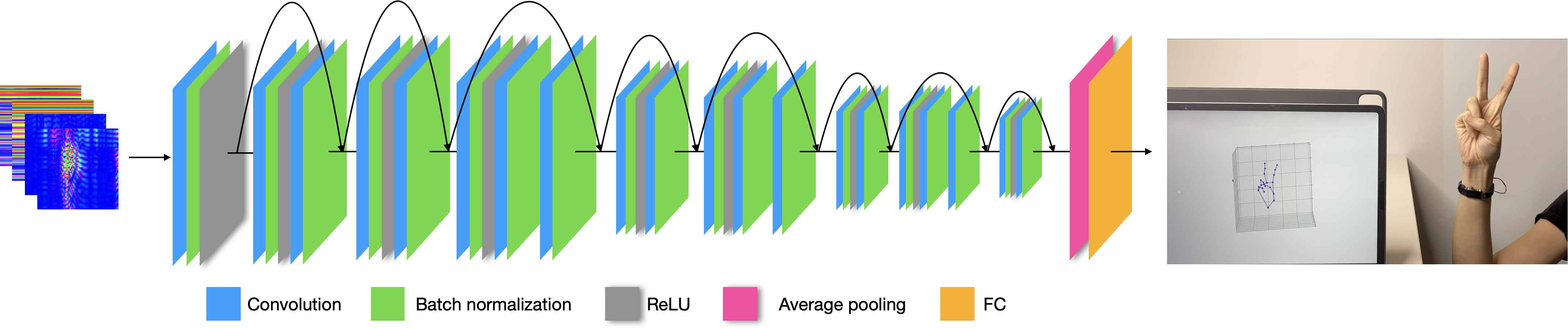}
  \caption{The model architecture of \name{}.}
  \Description{The model architecture of \name{}.}
  \label{fig:model}
  \vspace *{-1.2em}
\end{figure*}

\subsubsection{Echo profile analysis}
We employed the FMCW-based echo profile analysis as the sensing method, which has demonstrated promise in previous works~\cite{cfmcw, li2022eario, echospeech}. With a sampling rate of 50kHz and a frequency range of 20k-24kHz, chosen to be inaudible for humans, the signals of one frequency sweep are denoted as one \textit{FMCW frame}. To eliminate other frequencies, we applied a 20-24kHz bandpass filter. Subsequently, cross-correlation between transmitted and received signals was performed to obtain \textit{echo profiles}, which are formed by temporally stacking \textit{echo frames} and represent the reflection strength of signals traveling from paths with certain distances. Using the current echo profile to subtract the previous one produces \textit{differential echo profiles}, which eliminate static reflections and focus on moving objects. Figure~\ref{fig:hand-gestures} illustrates various hand gestures and their corresponding echo profiles.
Both original and differential echo profiles were employed to capture both movements and the static status of the hand and surroundings. The echo profiles were cropped to concentrate on distances of interest. For hand tracking, we used 72 pixels (24.6 cm) to focus on the hand, and for hand-object interaction, we used 88 pixels (30.2 cm) to encompass the hand's surroundings.

\subsubsection{Deep Learning Inference} \label{sec:deep-learning}
After obtaining the echo profiles, the information related to the hand postures is represented by a 2D image-like array. Due to its wide application and success in image processing, we used a customized deep Convolutional Neural Network (CNN) model to infer the 3D hand shapes or hand-object interactions. We stacked the original and differential echo profiles in channels. For hand pose tracking, we employed a shorter window length of 72 (0.864s) to predict the hand shape at the last moment of the window. For hand-object interaction, we incorporated a greater temporal context by employing a longer window length of 1050 (12.6s) since the activities are more intricate. In our pilot study with the researchers, we observed significant variability in the time taken to complete each hand-object interaction. To accommodate this variability and prevent information loss, a larger window was utilized. This resulted in input sizes of $72\times72\times4$ and $1050\times88\times4$, respectively.

The model architecture incorporates a ResNet-18 backbone, followed by an average pooling layer, a dropout layer (with a dropout rate of 0.8), and a fully connected layer. In the task of recovering the 3D hand shape, the output shape is 60, corresponding to 20 3-D coordinates (without the wrist, which serves as the origin). The wrist is not included in this output as it serves as the origin. In the case of wrist rotation recovery, the output shape is 3, representing the wrist-to-palm vector. To prioritize the joints with larger errors, the model employs Mean Squared Error (MSE) loss. During the training process, Adam's optimizer is utilized, with an initial learning rate set to 0.0002. The batch size for training is set to 30.

In the classification task of hand-object interaction, a similar architecture is employed, although the final layer of the CNN is replaced with a linear classifier. For this task, the output shape is 12, signifying the number of classes, and Cross-Entropy (CE) loss is employed to facilitate accurate classification.

\subsubsection{Data augmentation}
We applied data augmentation to improve the robustness of the model. During training, the echo profiles were randomly vertically shifted by $\pm11$ pixels. This was applied to compensate for the variance caused after remounting, where the position of the wristband may shift vertically. In 80\% of cases during training, each pixel in the echo profiles was multiplied by a random factor between 0.95 and 1.05. This is applied to avoid overfitting to a fixed set of values after multiple epochs.

\subsection{Evaluation Metrics} \label{sec:evaluation-metric}
While the assessment of Simple Gestures and Complex Gestures primarily centered on joint movements, the evaluation of Wrist Orientations specifically focused on wrist rotation. As a result, we employed different metrics for the two tasks:
\subsubsection{3D Hand Pose Estimation} To gauge the precision of our 3D hand pose estimation, we employed two established metrics: mean joint Euclidean distance error (MJEDE) and mean joint angular error (MJAE). These metrics have been utilized in prior works~\cite{hu2020fingertrak, kim2022ether, liu2021wr, kim2012digits, zhou2022learning}. 

The MJEDE of each frame is calculated by averaging the Euclidean distance of 20 joints (excluding the wrist).
% :
% $$\mathrm{MJEDE} = \frac{1}{20}\sum_{j=1}^{20}\sqrt{(x_j - \hat{x_j})^2 + (y_j - \hat{y_j})^2 + (z_j - \hat{z_j})^2}$$
% , where $(\hat{x_j}, \hat{y_j}, \hat{z_j})$ is the 3D coordinate of the ground truth for joint $j$.
The MJAE is calculated by averaging the angular error of the 15 joint angles (excluding five fingertips). Each joint angle is the angle between two consecutive bone segments.

In addition to MJEDE and MJAE, we also present the error distribution and the error of each joint or joint segment.

\subsubsection{Wrist Rotation Estimation}
To evaluate the performance of \name{} on wrist rotation estimation, we use the mean wrist angular error (MWAE) as the evaluation metrics, which is calculated by the angular error of the wrist-to-palm vector. In this context, the wrist-to-palm vector $v_9$ is defined as the vector pointing from the wrist to the metacarpophalangeal joint of the middle finger (Fig.~\ref{fig:mediapipe}(b)). Similarly, we reported the error distribution.
% For each frame, we use the following formula to calculate the the wrist angular error: 

% $$\mathrm{WAE} = \mathrm{ang}<v_9, \hat{v_9}> = \mathrm{arccos}\frac{v_9 \cdot \hat{v_9}}{||v_9|| \cdot ||\hat{v_9}||}$$

% \subsection{Echo Profile Calculation} \label{sec:echo-profile}
% \rz{maybe remove this for sake of length}
% \subsection{Data Collection Interface}
% \rz{maybe move this to study and only briefly mention}
% For the ease of the data collection process,
% % While conducting the gesture and activity recognition portion of the study, 
% we designed a graphical interface to show participants instructions on the screen for them to perform while they were being recorded through the webcam. By synchronizing the webcam video and instruction intervals with the data collected by the device, we attain the labels that are associated with participants' actions. Further detail will be described in Section~\ref{sec:evaluation-2}.

\section{Evaluation Overview}
We intend to build a wristband that can not only understand the hands themselves but also the objects that they interact with. For this purpose, we designed two studies to assess the feasibility of using  \name{}'s in continuous 3D hand pose tracking and hand-object interaction recognition, respectively. The studies were approved by the institution review board (IRB) of our institution.

With the first study, we designed three sets of hand gestures to demonstrate that \name{} is able to continuously recover the 3D hand pose when there is no object in the hand. With the second study, we incorporated 12 activities to demonstrate that beyond the tracking of free-hand postures, \name{} possesses the ability to comprehend hand-object interactions. We introduce the details of study design, procedures, and performance in the following sections.

% For the first study, we designed 3 sets of hand gestures
\section{User Study 1 - Hand Pose Tracking} \label{sec:evaluation-1}

\begin{figure*}[h]
  \centering
  \includegraphics[width=2\columnwidth]{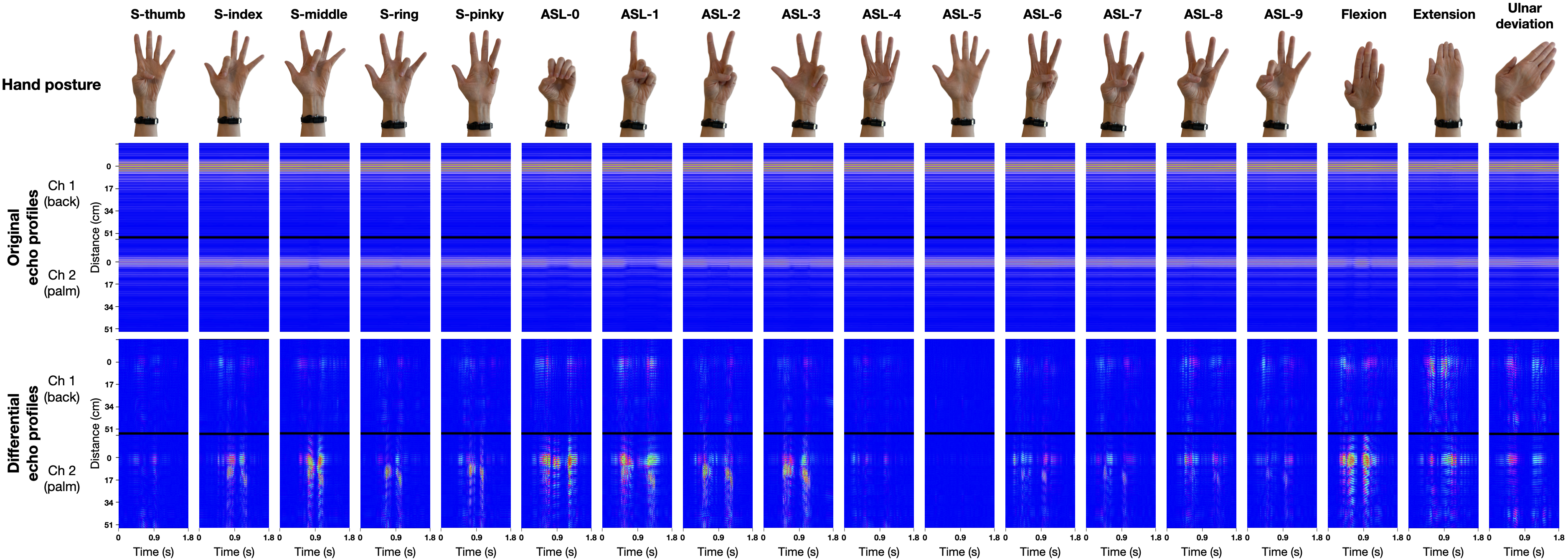}
  \caption{Illustration of gestures used in hand pose tracking and their echo profiles. 18 gestures in 3 categories: simple gestures (gestures stating with S-), complex gestures (10 American Sign Language gestures, starting with ASL-), and 3 wrist rotations (the last three columns).}
  \Description{Photos and echo profiles of 18 hand poses. The poses are: S-thumb, S-index, S-middle, S-ring, S-pinky, ASL-0 to ASL-9, Flexion, Extension and Ulnar deviation. Echo profiles include original echo profiles and different profiles. Original echo profiles are mostly blue, with thick horizontal strips. They are similar across all gestures. The differential echo profiles are also mostly blue, but have irregular color pixels clustered together in different areas. Distinct patterns can be observed for different gestures.}
  \label{fig:hand-gestures}
  \vspace *{-1.2em}
\end{figure*}

% A user study was conducted to evaluate the performance of \name{} in estimating 3D hand pose and wrist orientation. The study was approved by the institution review board (IRB) of our institution.

% \subsection{Study Design}
The objective of this first study was to evaluate the performance of \name{} in tracking 3D hand poses under different conditions. First, the study analyzed the device's accuracy in tracking different finger gestures across multiple sessions after the device was remounted. Second, the study assessed the effectiveness of the results obtained from training on a different number of sessions of data from each participant.

To explore the efficacy of hand pose tracking with different levels of complexity, three gesture sets were used in the study. The first two sets were designed to confirm the accuracy of 3D hand pose tracking, while the third set was specifically designed to validate the accuracy of wrist rotation estimation:

\textbf{(1) Simple Gestures:} This set of gestures consisted of five single-finger movements, wherein only one finger would bend at a time (Fig.~\ref{fig:hand-gestures} (a)). The purpose of these gestures was to assess the precision of \name{} in tracking and distinguishing between distinct finger postures.

\textbf{(2) Complex Gestures:} This set of gestures included ten complex finger gestures, modeled after the American Sign Language (ASL) finger gestures representing the digits 0-9 (Fig.~\ref{fig:hand-gestures} (b)). Unlike Simple Gestures, these gestures involve fingers occluding each other, introducing a higher level of complexity for tracking and recognition. Notably, this particular set of gestures was previously utilized in a prior work~\cite{kim2022ether}, explicitly selected to challenge \name{}'s capability to accurately track more complicated motions with multiple finger movements and occlusions.

\textbf{(3) Wrist Orientations:} This set of gestures comprised three distinct wrist orientations: flexion, extension, and ulnar deviation  (Fig.~\ref{fig:hand-gestures} (c)). It is noteworthy that radial deviation was excluded due to its reported difficulty in the pilot study. These wrist motions are essential in everyday hand gestures. As \name{} is mounted on the wrist, it has the unique advantage of observing the hand from that vantage point, making it possible to track wrist orientations accurately.

\subsection{Participants}
We recruited 12 participants (4 self-reported males, 8 females) aged 20-26 (M = 22.0, SD = 1.5) with snowball sampling at a local university. Ten participants self-identified as right-handed, two left-handed. However, all participants reported wearing or intending to wear watches or wristbands on their left wrists. Consequently, all participants wore the device on their left wrist during the study. After the study, the participants were compensated \$15 in local currency.

\subsection{Apparatus}
The study was conducted in a quiet experiment room. Participants were seated at a table with a laptop positioned in front of them, serving as a platform for presenting instructions. For capturing the ground truth data, a webcam~\footnote{Logitech C615~\url{https://www.logitech.com/en-us/products/webcams/c615-webcam.960-000733.html}} was employed to record the movements of the participant's hand. These recorded videos of hand posture were processed by MediaPipe~\cite{lugaresi2019mediapipe} to extract the ground truth, including the 3D positions of 21 finger joints. As previously mentioned, all participants wore the device on their left wrist, and a sticker was used to mark the precise position for ease of remounting. As the wrist thickness varied among people, the device was customized by adjusting the distance between the two pairs of speakers and microphones.

During the user study, we streamed the acoustic data from the two microphones to a smartphone (Redmi Note 10 Pro) via Bluetooth Low Energy (BLE). However, due to heavy traffic, we occasionally experienced packet loss during transmission. In such cases, the lost packets were replaced with zeros, and the data containing the lost periods was removed from the dataset. Throughout the entire study, we experienced only a 0.35\% packet loss in BLE.

\subsection{Procedure}
The procedure followed in the user study was as follows:

\textbf{(1) Introduction:} The study started with participants signing a consent form and receiving an introduction to the study's procedure. 

\textbf{(2) Practice Session:} The study was divided into 21 data collection sessions, each following the same process. The initial session served as a practice session, allowing the participants to get familiarized with the testing system and the act of performing gestures. Note that the data from the practice session was used for neither training nor testing later. The participants were not informed of this distinction and treated the practice session as any formal session.

\textbf{(3) Data Synchronization:} During each session, the researcher would snap their fingers in front of the camera to mark the beginning and end of the session. The sound of the snap was captured by the microphones on the device, while the finger gesture was recorded by the camera. This allowed for the synchronization of the acoustic data and the ground truth video footage. 

\textbf{(4) Data Collection:} In each session, the aforementioned three gesture sets were presented in the following order: (1) Simple Gestures, (2) Complex Gestures, and (3) Wrist Orientations. Within each gesture set, each gesture, lasting 2 seconds, was repeated four times. The sequence of gestures was randomized to mitigate potential learning effects.

We created a user-friendly graphical interface to present instructions and streamline the user study process. The interface was displayed on the laptop screen. At the beginning of each trial, an image representing the target gesture appeared, and the participant had 2 seconds to replicate the gesture. The interface also featured a countdown timer for the ongoing gesture and displayed the number of remaining trials. The participants were explicitly instructed to mimic the gesture displayed on the screen and return to a neutral position upon completing each gesture.

\textbf{(5) Device Remounting:} After each session, the participant was asked to take a short break and then remount the device before the start of the next session. Participants had the freedom to relax their arms or move around during these breaks. This remounting procedure was designed to evaluate the performance of our system in real-world scenarios where users frequently remove and reattach the device. 

Each session lasted approximately 2.5 minutes. Accounting for the intervals between sessions, the entire study extended for a duration of 75 to 90 minutes for each participant. In total, each participant contributed 1440 (= (5 (Simple Gestures) + 10 (Complex Gestures) + 3 (Wrist Orientations)) $\times$ 4 (repetitions) $\times$ 20 (sessions excluding the practice session)) gestures to the dataset, which means that 17280 gestures were collected.

It is worth mentioning that some participants occasionally performed incorrect gestures that differed from the target gestures. However, we decided to include this data in our dataset since the video ground truth faithfully captured the actual gestures performed by the participants. Additionally, many participants experienced physical limitations that made it challenging for them to execute certain hand gestures, such as being unable to bend certain fingers without bending others. In such cases, the participants were instructed to mimic the gestures in a way that was comfortable to them. As a result, our dataset contains many non-standard instances of gesture.

\subsection{Training Scheme} \label{sec:training-scheme}
To minimize training effort from a new user, we seek to maximize the utilization of data collected from other users. Therefore, we developed a two-step training-fine-tuning scheme. In the first step, we trained a model using data collected from other users. For each new user, we fine-tune the pre-trained model with only a small set of training data collected on the new user. Compared with training a user-dependent model from scratch directly on data provided by this new user, the model not only converges faster but also yields better performance. 

For the evaluation, we first trained a leave-one-participant-out (LOPO) model for each participant. Please note that this is a user-independent (UI) model. The UI model was trained for 10 epochs. We then fine-tune this UI model with different amounts of data collected from the same participant for another 5 epochs. The same learning rate (0.0002) was used in both steps.

\subsection{Results - 3D Hand Pose Estimation} \label{sec:gesture-performance}
Following the two-step training scheme (Section~\ref{sec:training-scheme}), we first trained a LOPO model for each participant and then fine-tuned the model with the specific participant's data. The fine-tuned model was tested on the last two sessions from the participant's data. When we used the first 18 sessions (i.e., all sessions except the testing ones) as training data to fine-tune the LOPO model, our results yielded an MJEDE of 4.81 mm (SD = 0.99 mm) and MJAE of 3.79\textdegree (SD = 0.68\textdegree) across all participants. It's noteworthy that this performance remained consistent among participants. 

\begin{figure*}[h]
  \centering
  \includegraphics[width=2\columnwidth]{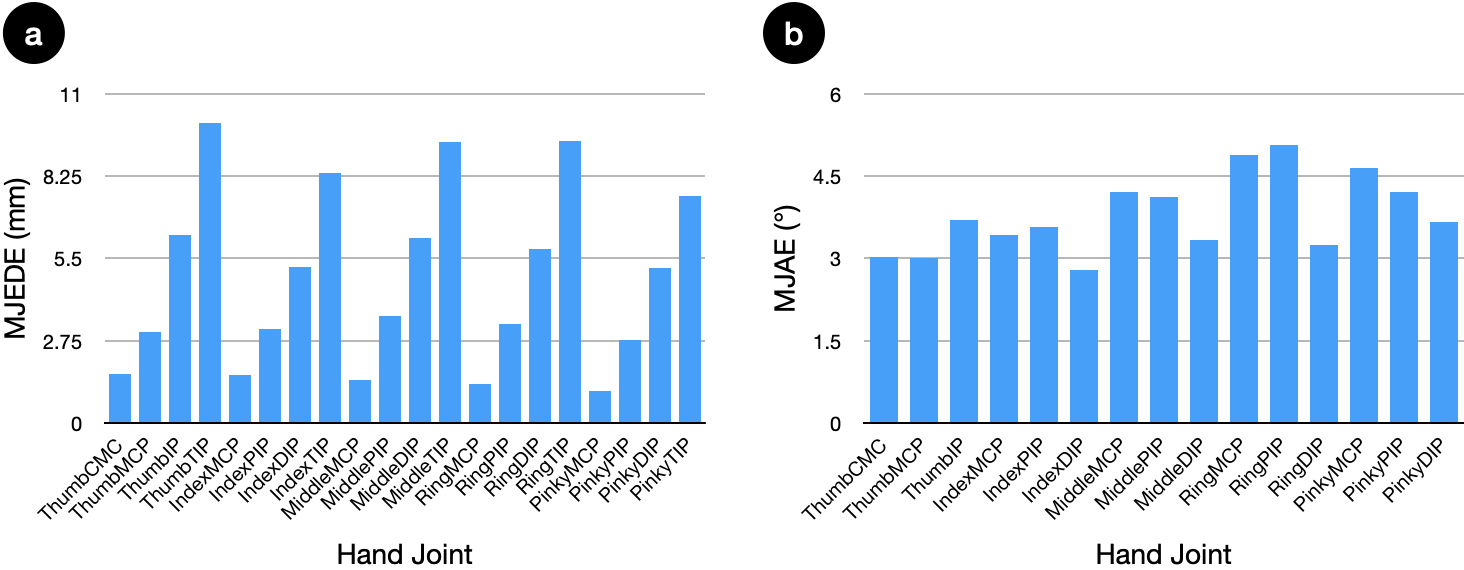}
  \caption{3D hand shape reconstruction performance on all hand joints. (a) Using MJEDE metric. (b) Using MJAE metric.}
  \Description{(a) Bar chart. H axis: ThumbCMC, ThumbMCP, ThubmIP, ThumbTIP, IndexMCP, IndexPIP, IndexDIP, IndexTIP, MiddleMCP, MiddlePIP, MiddleDIP, MiddleTIP, RingMCP, RingPIP, RingDIP, RingTIP, PinkyMCP, PinkyPIP, PinkyDIP, PinkyTIP. V axis: MJEDE (mm). Values: 1.621	3.028	6.275	10.008	1.612	3.136	5.217	8.351	1.425	3.569	6.166	9.389	1.29	3.306	5.815	9.421	1.052	2.781	5.193	7.587. (b) Bar chart with the same H axis. V axis: MJAE(\textdegree). Values: 3.025	3.013	3.694	3.426	3.576	2.789	4.214	4.124	3.342	4.882	5.071	3.235	4.639	4.209	3.662.}
  \label{fig:joint-performance}
  \vspace *{-1.2em}
\end{figure*}

\begin{figure*}[h]
  \centering
  \includegraphics[width=2\columnwidth]{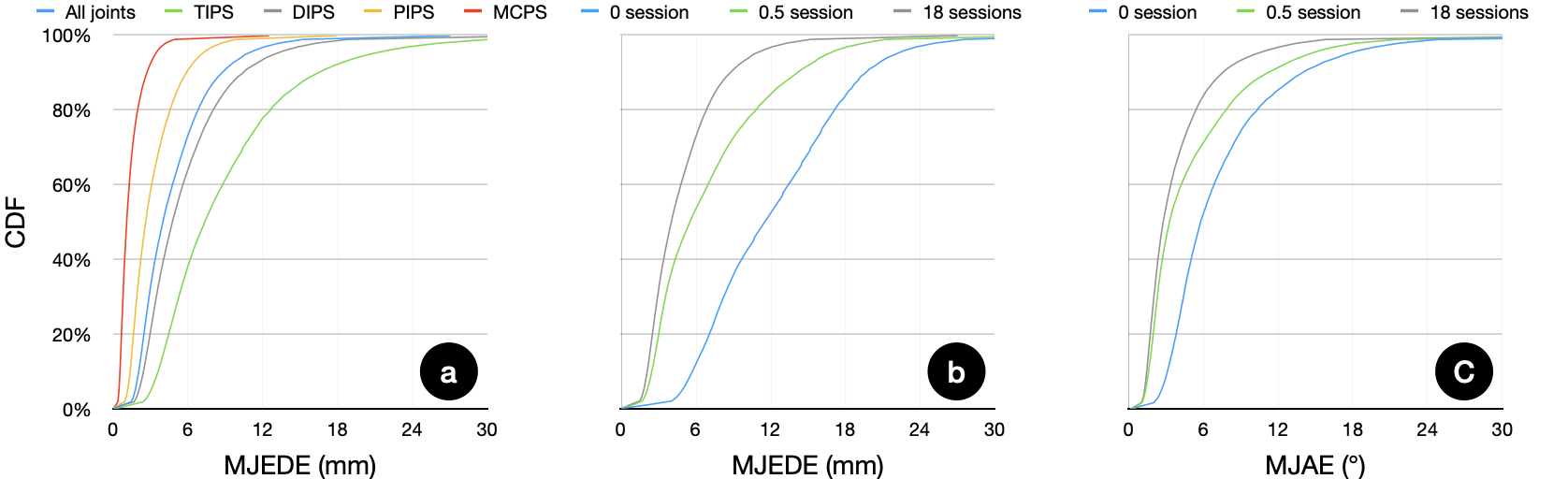}
  \caption{Error distribution in 3D hand shape reconstruction. (a) The error distribution of MJEDE of different joint types. (b) The error distribution of MJEDE of all joints when 0, 0.5, 18 sessions of data was used to fine-tune the UI model. (c) The error distribution of MJAE of all joints when 0, 0.5, 18 sessions of data was used to fine-tune the UI model.}
  \Description{All CDF functions. (a) Five convex curves, MCPS goes to 100\% fastest, followed by PIPS, All joints, DIPS and TIPS. H axis most under 18mm MJEDE. (b). 3 curves, 18 sessions goes to 100\% fastest followed by 0.5 sessions and 0 session. H axis mostly under 24mm MJEDE. (c) Same as (b) but H axis is MJAE, mostly under 24 \textdegree.}
  \label{fig:cdf}
  \vspace *{-1.2em}
\end{figure*}

In general, fingertips exhibited the largest error, with an average of 10 mm, across all finger joints. Among the five fingertips, the thumb's fingertip had the highest error, measuring 10.0 mm. However, this discrepancy was not significantly larger than that of other fingertips. The joint angle at the proximal joint of the ring finger has the largest angular error of 5.1\textdegree (Fig.~\ref{fig:joint-performance}). Specifically, although fingertips tend to have larger Euclidean distance errors, the angular errors of all joints are similar. This indicates that the larger error on the tips of the fingers largely comes from accumulated error from connected joints. The error distribution of MJEDE of the finger TIPs, DIPs, PIPs, and MCPs are illustrated in Fig.~\ref{fig:cdf}(a). Performance on single-finger and complex-finger gestures are similar: the MJEDE on simple and complex gestures is 4.69 mm (SD = 1.17 mm) and 4.87 mm (SD = 0.94 mm), while the MJAE is 3.64\textdegree (SD = 0.75\textdegree) and 3.87\textdegree (SD = 0.67\textdegree). This shows that \name{} works consistently in estimating the hand poses while performing hand poses with different complexities.

\begin{figure*}[h]
  \centering
  \includegraphics[width=2\columnwidth]{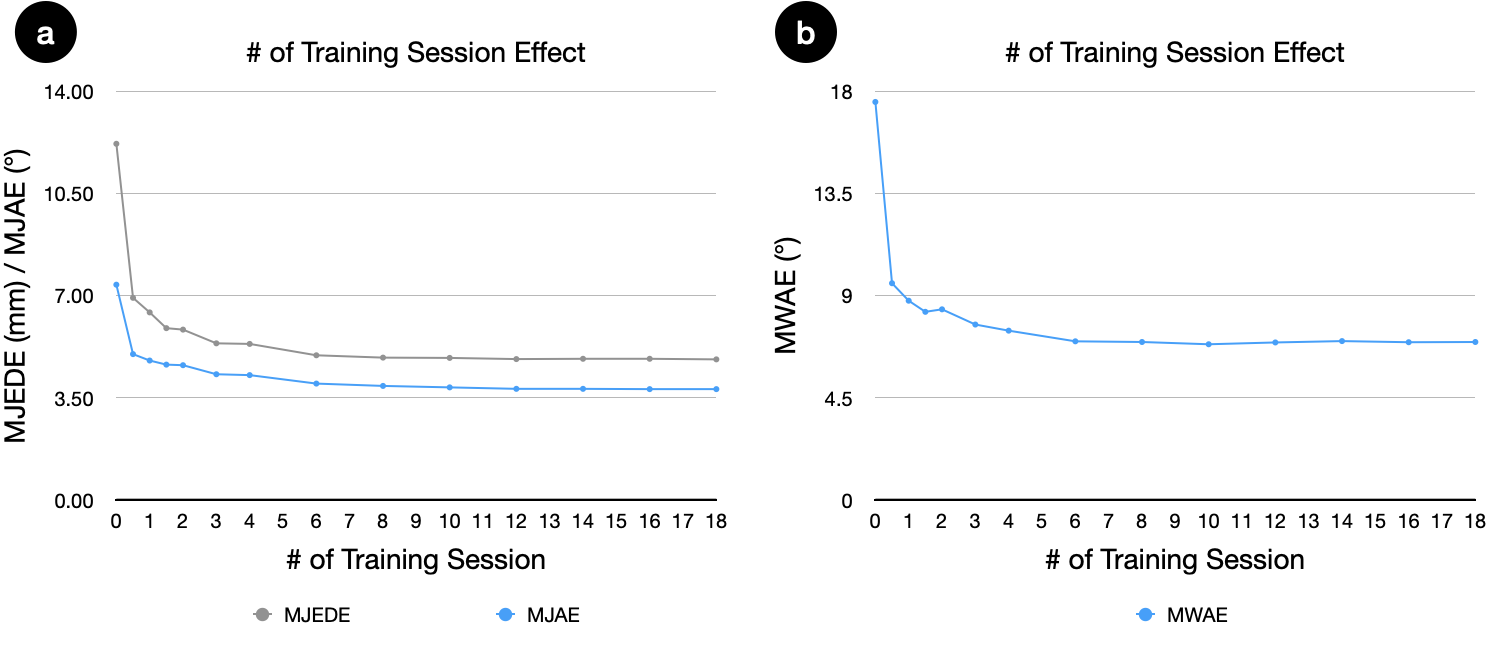}
  \caption{Adjusting the number of sessions used during the fine-tuning step. (a) Performance in 3D hand shape reconstruction. (b) Performance in wrist rotation recovery.}
  \Description{(a) Line chart, 2 series. MJEDE and MJAE. H axis: 0, 0.5, 1, 1.5, 2, 3, 4, 6, 8, 10, 12, 14, 16, 18. Values: MJEDE: 12.20 6.92, 6.42, 5.88, 5.83, 5.36, 5.34, 4.95, 4.87, 4.86, 4.82, 4.83, 4.83, 4.81. MJAE: 7.37, 4.99, 4.77, 4.63, 4.61, 4.30, 4.27, 3.98, 3.90, 3.85, 3.80, 3.80, 3.79, 3.79. (b) Line chart, 1 series. Same H axis, series MWAE: 17.53, 9.54, 8.77, 8.28, 8.39, 7.72, 7.45, 6.98, 6.95, 6.85, 6.93, 6.99, 6.94, 6.95.}
  \label{fig:pose-session}
  \vspace *{-1.2em}
\end{figure*}

The results from the above experiments showed promising tracking performance. However, it required 18 sessions of training data from each participant, which took 36 minutes to collect. It may not be preferred by many users if a new user has to collect such a long period of training data. The high demand for training data has been a long-lasting problem for the data-driven sensing approach. 

Therefore, in the follow-up experiment, we examined how much training data is actually needed for a new participant without significantly compromising the performance. To do this, we manipulated the number of sessions used during the fine-tuning stage. When no data from the same participant is used, the MJEDE and MJAE across all participants are 12.2 mm (SD = 3.73mm) and 7.37\textdegree (SD = 1.73\textdegree), respectively. Note that this is performance in the user-independent experiment of \name{}. When only using 0.5 sessions (about 1 minute) of data, the performance improved to MJEDE 6.92 mm (SD = 2.42 mm) and MJAE 4.99\textdegree (SD = 1.28\textdegree)(Fig.~\ref{fig:pose-session} (a)). The error distribution when 0, 0.5, and 18 sessions were used for fine-tuning is presented in Fig.~\ref{fig:cdf} (b, c). As the figure shows, the performance improves with more training data. However, we noticed the performance stopped increasing with 8 sessions of training data (20 minutes). This indicates that a new user only needs to provide 20 minutes of training data to obtain optimized performance. Even with only 1 minute of training data, \name{} can still achieve decent performance. The results from this study were very encouraging, which shows the potential of using the proposed sensing technology for real-world users with minimal calibration from each new user.

\subsection{Results - Wrist Rotation Estimation} \label{sec:wrist-performance}

\begin{figure*}[h]
  \centering
  \includegraphics[width=2\columnwidth]{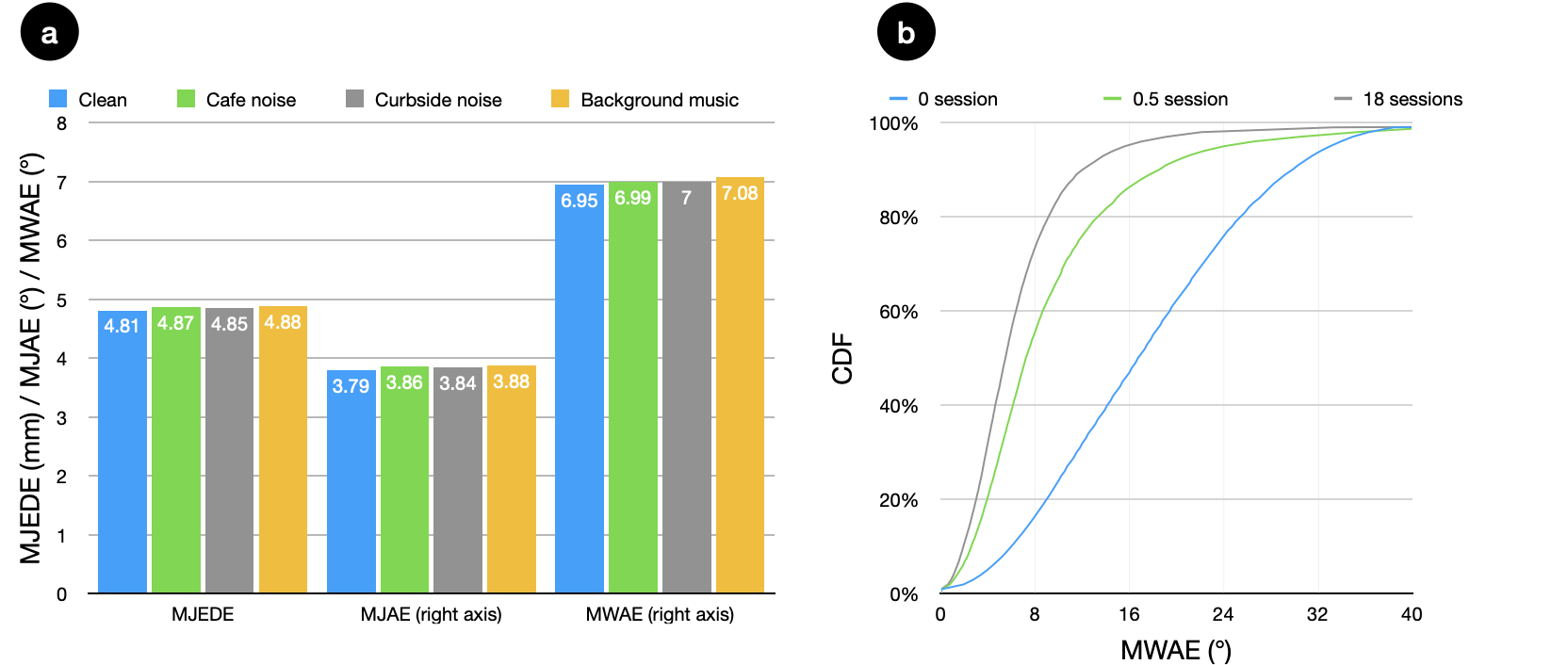}
  \caption{(a) Performance of \name{} with different injected noises. (b) Error distribution when 0, 0.5, 18 sessions from the same user was used to fine-tune the UI model.}
  \Description{(a) Bar chart with 4 series: Clean, Cafe noise, Curbside noise Background music. H axis: MJEDE: 4.81, 4.87, 4.85, 4.88. MJAE: 3.79, 3.86, 3.84, 3.88. MWAE: 6.95, 6.99, 7, 7.08. (b) CDF with 3 series. 0, 0.5 and 18 sessions. 18 goes to 100\% fastest, followed by 0.5 and 0. H axis MWAE mostly under 32.}
  \label{fig:noise}
  \vspace *{-1.2em}
\end{figure*}

We follow the same approach as described in Section~\ref{sec:gesture-performance} to evaluate \name{}'s performance in estimating wrist rotation. When using 18 sessions for fine-tuning, the mean wrist angular error (MWAE) across all participants is 6.95\textdegree (SD = 2.23\textdegree). The error distribution of MWAE is demonstrated in Fig.~\ref{fig:noise}(b). By adjusting the number of sessions used during fine-tuning, we obtain the curve shown in Fig.~\ref{fig:pose-session}(b). In the User-Independent (UI) setting, where no training data from the same participant was used, the performance is 17.5\textdegree (SD = 5.75\textdegree). With only 0.5 sessions (72 seconds) of training data, \name{} still achieves a performance of 9.54\textdegree. As a special note on P02, whose wrist rotation estimation performance was significantly worse than others, we analyzed the recorded video and found that the ground truth of P02's hand pose was not stable or accurate. P02 performed the flexion gesture in close proximity to the camera with a large angle, which caused the camera to easily lose focus and MediaPipe to struggle with capturing the hand reliably due to the blurry image and skewed view angle. In later studies, the camera was placed further away to enable participants to perform gestures without significantly impacting image quality.

\subsection{Results - Noise Injection} \label{sec:noise}
\name{} uses active acoustic sensing as the sensing principle. The frequency range of the signals used during the study is 18-21 kHz, which is well beyond the range of human conversations and most noises. However, in order to examine \name{}'s robustness against various acoustic noises existing in the real world, we recorded noises in three different scenarios as specified in Table~\ref{tab:noisescenarios} and injected the noises into the testing data. Please note that the model was trained using the training data without noise injection. In this setting, we can evaluate how our system would react to different noises without the need to collect training data for each noise. 

\begin{table*}[ht]
    \centering
    \caption{Noises recorded in different scenarios.}
    \Description{Table. First row: Scenario, Cafe, Curbside, Background music playing. Second row: Noise level (dB (A)) (with footnote 5), 61.8, 71.5, 70.4.}
    \begin{tabular}{|c||c|c|c|}
    \hline
        Scenario & Cafe & Curbside & Background music playing \\
        \hline
        Noise level (dB (A)) \tablefootnote{Measured by NIOSH Sound Level Meter App: \url{https://www.cdc.gov/niosh/topics/noise/app.html}} & 61.8 & 71.5 & 70.4 \\
        \hline
    \end{tabular}
    \label{tab:noisescenarios}
\end{table*}

Results (Fig.~\ref{fig:noise}) indicate that there is nearly no performance degradation when different types of noises are injected. The maximum performance drop is on background music playing, where MJEDE increases from 4.81 mm to 4.88 mm. We performed one-way ANOVA tests on all three noises and did not find a significant difference between performance in the clean environment and with different noise injections: for 3D hand pose estimation, F(3, 44) = 0.010, p = 1.00 > 0.05; for wrist rotation estimation, F(3, 44) = 0.0066, p = 1.00 > 0.05.

\subsection{Follow-Up Study - Hand Motion Speed}
As \name{} relies on both original and differential echo profiles, we want to investigate whether the hand motion speed affects the tracking performance. To address this, we conducted a follow-up study, collecting data from participants executing the same gestures at different speeds.

\subsubsection{Participants} We recruited another 12 participants (5 self-reported males, 7 females) aged 18-31 (M = 24.1, SD = 4.8) using snowball sampling at a local university. Eleven participants self-identified as right-handed, one left-handed. To align with the initial study, we asked all the participants to wear the device on their left wrist during the study. After the study, the participants were compensated \$20 in local currency.

\subsubsection{Apparatus} The apparatus is closely identical to the initial study, with two differences. Firstly, a laptop's built-in camera~\footnote{MacBook Pro 14-inch, 2021~\url{https://www.apple.com/macbook-pro/}} was employed instead of the webcam. Secondly, the data was stored on a microSD card to prevent packet loss. This change in data storage is to avoid data loss since we collect less data for each condition in this study.

\subsubsection{Procedure} The procedure closely mirrors that of the initial study, involving 24 data collection sessions, each adhering to the same protocol. The participants were instructed to perform the gestures at varying speeds during each session. Three different speeds were implemented: (1) Fast (1.5 s/gesture), (2) Medium (2.0 s/gesture), and (3) Slow (2.5 s/gesture). Instead of images, videos showing a hand executing the gestures at the specified speed were shown to guide the participants, who were instructed to try their best to follow the movement/speed of the videos. The initial three sessions, one for each speed, were designated as practice sessions, and the data collected during this period was excluded from both training and testing later. The sequence of speeds was randomized for the remaining 21 sessions.

The entire study took 90 minutes for each participant. In total, each participant contributed 504 (= 18 (gestures) $\times$ 4 (repetitions) $\times$ 7 (sessions excluding the practice session)) gestures to the dataset for each of the three speeds, resulting in 3 datasets, each containing 6048 gestures. Note that one of the participants had an emergency during the study, so they completed the last five sessions the next day. In addition, due to a hardware issue, some data were lost for three participants. To address this, two participants were invited to redo one session on the other day, while the other one redid two. Each participant was compensated an additional \$15 in local currency.

\subsubsection{Training Scheme} The training scheme is the same as the initial study. A model was first trained for each participant at the Medium speed, consistent with our initial study, where each gesture was performed in a two-second interval. The User-independent (UI) model underwent training for 20 epochs. Subsequently, we fine-tuned the UI model using the data from a specific participant at the Medium speed for an additional 10 epochs, with a consistent learning rate of 0.0002. 6 sessions were utilized as training data for fine-tuning, while the remaining session served as the testing data. Following the fine-tuning on the Medium speed model, we tested the model using the 7 sessions of data at both Fast and Slow speeds from the same participant, respectively.

\subsubsection{Results} When we tested the model with the same, i.e., the Medium, speed, the average MJEDE is 11.01 mm (SD = 4.99 mm), and MJAE is 7.38\textdegree (SD = 1.89\textdegree). For testing the model with the Fast speed data, the average MJEDE is 12.26 mm (SD = 7.80 mm), and MJAE is 7.80\textdegree (SD = 2.09\textdegree). In the case of the Slow speed data, the average MJEDE is 13.47 mm (SD = 8.27 mm), and MJAE is 8.54\textdegree (SD = 2.50\textdegree). Note that the dataset used in this study was only one-third of the initial dataset, explaining the difference in performance compared to the initial study. In addition, the use of video instructions, as opposed to image instructions, led to shorter reaction times and posed challenges for some participants in following the instructions accurately. Despite instructing participants to complete the incorrect gestures, many attempted corrections, introducing more unseen poses and reducing the number of target poses in the dataset. Notably, one participant had a significantly larger hand, with the longest part measuring 22.8 cm, compared to the average of 18.2 cm for other participants. This variation contributed to unexpected outcomes. Nevertheless, the results of this study remain comparable to prior work \cite{devrio2022discoband, kim2022ether}.

% In our initial study, each gesture was performed at a two-second interval, equivalent to the Medium speed in this follow-up study. Consequently, our base model is primarily trained on Medium speed data. Thus, our emphasis was on comparing Medium speed data with that of Slow or Fast speed. For each participant, we fine-tuned the model using six sessions of Medium-speed data and tested it on the remaining session at the same speed. Furthermore, we fine-tuned the model with the same six sessions of data and tested it on data from the same participant but at a different speed.

In this study, we aimed to investigate the impact of motion speed on the performance of our hand pose tracking system. We initiated our analysis with a one-way ANOVA. In the case of MJEDE, the results indicated no significant difference ($F_{2, 249} = 2.75, p = 0.07$) when testing the Medium-speed model on datasets with Fast, Medium, or Slow speeds. This suggests that the performance of our system, as measured by MJEDE, remains consistent across different motion speeds. However, when considering MJAE, the results revealed a significant difference ($F_{2, 249} = 6.12, p < 0.05$). To further explore these differences, we conducted a Tukey test on MJAE. The results of the Tukey test indicated that there is no significant difference in the testing pairs [Medium, Fast] and [Fast, Slow]. In contrast, a significant difference was observed in the testing pair [Medium, Slow].

This finding suggests that while there is consistency in MJEDE across different motion speeds, there are variations in MJAE. Specifically, the difference observed in [Medium, Slow] suggests that there may be an impact on angular accuracy when the model is trained on a specific motion speed and tested on another. In summary, our comprehensive statistical analyses reinforce the overall robustness of our system, highlighting its ability to maintain consistent performance in terms of MJEDE across different motion speeds. The observed variations in MJAE, particularly in the [Medium, Slow] pair, provide valuable insights into the system's behavior under different motion speed conditions.
\section{User Study 2 - Hand-Object Interaction Recognition} \label{sec:evaluation-2}

\begin{figure*}[h]
  \centering
  \includegraphics[width=2\columnwidth]{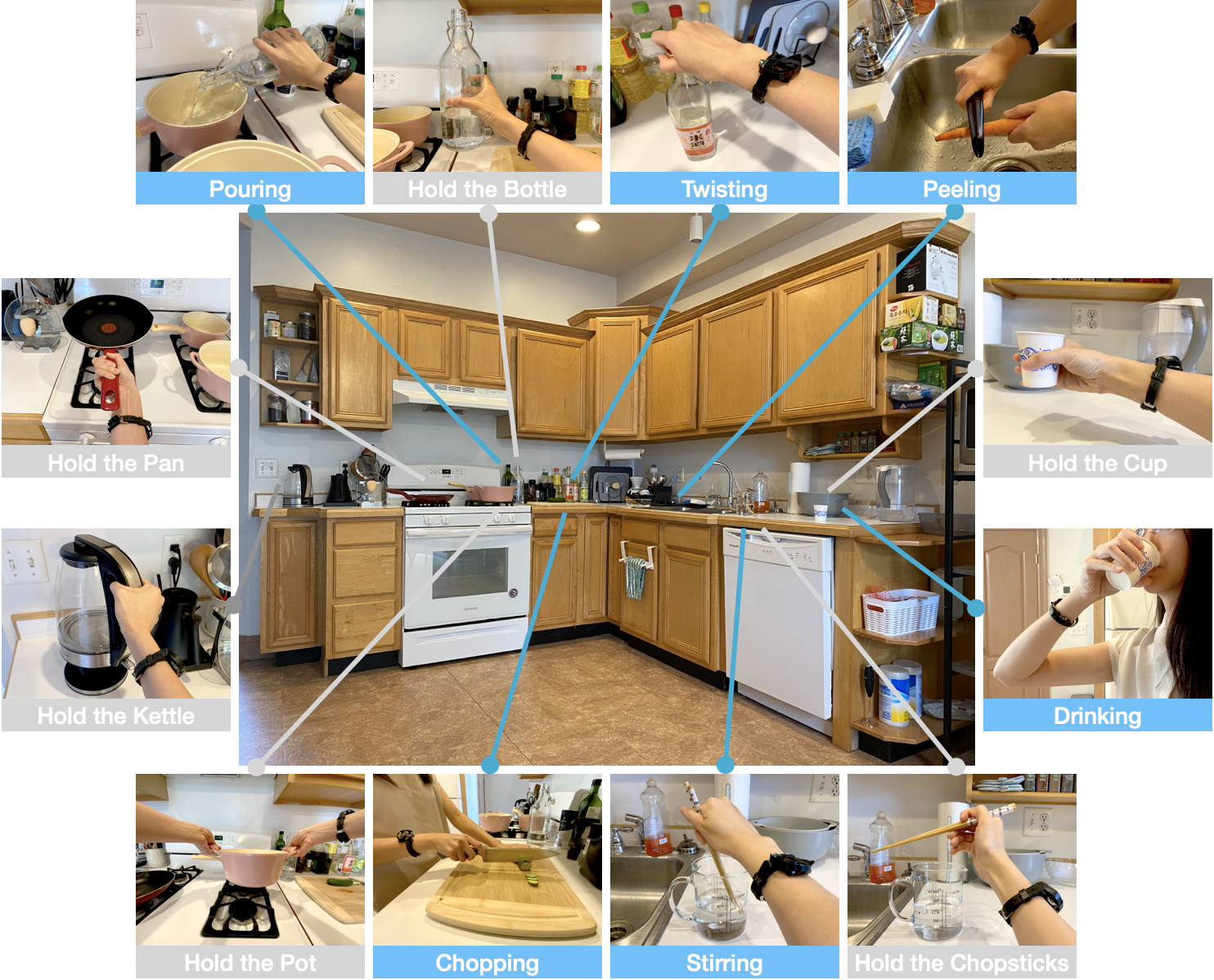}
  \caption{The kitchen and the activities used in the user study. Blue background indicates Dynamic Interactions while gray background indicates Static Interactions.}
  \Description{A photo of a kitchen at the center surrounding by 12 photos of hand-related activities. With blue backgrounds: Pouring, Twisting, Peeling, Drinking, Stirring and Chopping. With gray background: Hold the bottle, Hold the Cup, Hold the chopsticks, Hold the pot, Hold the kettle, Hold the bottle.}
  \label{fig:studystructure}
  \vspace *{-1.2em}
\end{figure*}

% Understanding the object a person is holding can provide crucial contextual information that unveils the user's specific and detailed activities. Another user study was conducted to measure the recognition accuracy of \name for everyday activities that involve hands in a real-life kitchen setting. The study was approved by the institution review board (IRB) of our institution.

% \subsection{Study Design}
In the second user study, our primary objective was to evaluate \name{}'s capability to recognize everyday hand activities within a naturalistic environment.
% Beyond the tracking of free-hand postures, as discussed in Section~\ref{sec:evaluation 1}, \name{} possesses the ability to comprehend hand-object interactions.
Owing to the active acoustic sensing techniques, \name{} can effectively recognize both \textit{Static Interactions}, such as holding objects in hand, and \textit{Dynamic Interactions}, such as moving objects. As a result, an interaction set, with half of them being static interactions and the remaining half being dynamic interactions, was used to comprehensively assess the device's performance in capturing these real-world activities.

On the other hand, although humans engage in extensive hand-object interactions across diverse contexts in our daily lives, we have confined our study's context to the kitchen for a proof of concept. This decision stems from the fact that the kitchen setting is a frequently chosen dataset scenario in prior research \cite{Damen2018EPICKITCHENS, Damen2022RESCALING, das2013thousand, zhou2018towards, rohrbach2012database}. It can be applied to support a wide range of applications, including but not limited to elder care, smart home technology, and accessibility solutions.

As a result, an interaction set consisting of 12 hand-object interactions specifically tailored to the kitchen environment was proposed. This included 6 \textit{Static Interactions}: \textbf{holding a paper cup}, \textbf{a pair of chopsticks}, \textbf{a glass water bottle}, \textbf{a pot}, \textbf{a pan}, and \textbf{a kettle}, and 6 \textit{dynamic interactions}: \textbf{drinking}, \textbf{stirring}, \textbf{peeling}, \textbf{twisting}, \textbf{chopping}, and \textbf{pouring}.

\subsection{Participants}
We recruited 12 participants (9 females, all right-handed) aged 21-33 (M = 27.0, SD = 3.7, one preferred not to state their age) with snowball sampling at a local university. The participants were compensated \$20 in local currency.

\subsection{Apparatus}
Different from the first study, we conducted this study mimicking real-life settings. Therefore, the participants conducted this study at one researcher's home (2 bedroom 2.5 bathroom townhouse), where their roommate continued with their usual activities, including generating background noise. The study took place in the kitchen area, which adjoined the living room. Participants moved around the kitchen to complete various tasks while the researcher remained in the living room to oversee the study. All objects, except for paper cups, were what the researcher regularly used at home and were in their usual place. Disposable paper cups were used for drinking to maintain hygiene standards. In addition, the participants had the freedom to engage in conversations with the researcher during the study.

In this study, the participants had to interact with different objects. As this interaction was mostly done through the dominant hand, they were requested to wear \name{} on their dominant arm (all the participants were right-handed), at the relative position where they usually wear a watch. To facilitate the remounting process, a sticker was used to mark the position. The device was customized for each participant based on the thickness of their wrist by adjusting the length between the two pairs of speakers and microphones. Additionally, the participants were asked to wear earbuds to allow for the use of voice commands throughout the study. We saved the data into microSD card to avoid unstable BLE data package loss with more complex electro-magnetic environment.

% In a real-world setting, numerous devices using BLE were present, including televisions, earbuds, smartphones, and laptops. This extensive usage of BLE in the vicinity could potentially compromise the stability of data streaming through BLE. Furthermore, considering that each session in this study lasted approximately 8 minutes, we wanted to mitigate any risk of packet loss. Therefore, unlike the first study, where BLE was employed for data collection, we opted to store the acoustic data directly onto the device's microSD card.

\subsection{Procedure}
The study shared similar procedures with the following differences:
% The procedure followed in the user study was as follows:
% \textbf{(1) Pre-Study Survey:} Before the study began, the participants were required to sign a consent form and complete a pre-study survey, which collected their demographic information. 

\textbf{(1) Introduction:} A demonstration of the 12 interactions and the environment configuration was included. Each participant was asked to go through the 12 interactions independently to confirm their full comprehension of the procedure and the setup.

% \textbf{(3) Data Synchronization:} In each session, the researcher would clap their hands in front of the camera to signal the start and end of the session. This was done for synchronization, mirroring the snap used in the first study. The clap was used in this study because the participant was in the kitchen, which was around two meters away from the researcher. The clap was chosen due to its greater volume and ease of detection by the device on the participant.

\textbf{(2) Data Collection:} The study consisted of 5 data collection sessions, all of which followed the same process. Given that all the interactions featured in our interaction set were everyday activities that should be inherently familiar to all participants, there was no need for practice sessions, unlike in the first study.

During each session, participants were engaged in a series of interactions, each lasting 10 seconds and repeated four times. To minimize any learning effects, the order of these interactions was randomized. 

Voice commands were used in this study to present instructions and streamline the user study process. At the beginning of each trial, a voice command regarding the target interaction was issued to the participant. Subsequently, participants were allotted a 10-second timeframe to execute the interaction as instructed.

For the \textit{Static Interactions}, the participants were instructed to hold the object until the next interaction was presented. For the \textit{Dynamic Interactions}, the participants were granted the freedom to perform the interaction as many times as they wished within the 10-second timeframe. If the interaction was completed before the 10 seconds elapsed, participants could take a brief rest while placing their hands in a comfortable position. 

% \textbf{(5) Device Remounting:} Following each session, participants were given the opportunity to take a break, during which they were required to reattach the device. This procedure mirrored the protocol followed in the first study. 

% \textbf{(3) Post-Study Survey:} After all the data collection sessions, the participants were asked to fill out a post-study survey on a 7-point Likert scale (1=strongly disagree and 7=strongly agree), asking about the experience of using \name{}.

Each session lasted about 8 minutes. When accounting for the intervals between sessions, the entire study spanned a duration of 75 to 90 minutes for each participant. In total, 2880 (= 12 (participants) $\times$ 12 (interactions) $\times$ 4 (repetitions) $\times$ 5 (sessions)) interactions were collected.

\subsection{Results}

\begin{figure*}[h]
  \centering
  \includegraphics[width=2\columnwidth]{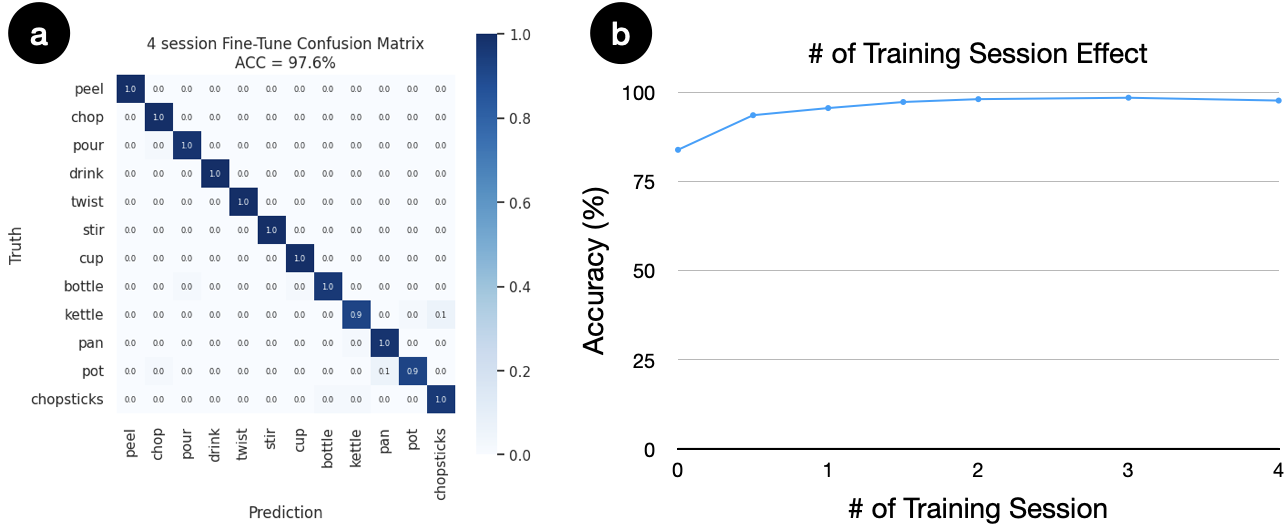}
  \caption{(a) The confusion matrix of 4-session fine-tune model. (b) The performance trend of adjusting the number of sessions used during the fine-tuning step.}
  \Description{(a) A confusion matrix. Title: 4 session Fine-Tune Confusion Matrix ACC=97.6\%. The diagnal series are almost all 1.0 except for kettle and pot. (b) Line chart. H axis: # of training session. V axis: Accuracy(\%). The series starts around 83\% and rises to over 90\% at 0.5, then flattens around 1.5 at around 95\%.}
  \label{fig:cm}
  \vspace *{-1.2em}
\end{figure*}

Following a similar two-step training scheme as described in Section~\ref{sec:training-scheme} used in the first study, we evalauted \name{}'s performance in recognizing hand-object interactions in both UI and UD ways. To do so, we first trained a LOPO model for each participant and then fine-tuning the model using only the specific participant's data. After fine-tuning the LOPO model with four sessions (i.e., all sessions except the testing ones) of data from the participant, the overall accuracy by averaging the results from all the participants was 97.6\% (SD = 0.82\%), and all the interactions have at least 80\% accuracy (Fig.~\ref{fig:cm} (a)). More specifically, all the Dynamic Interactions achieved 100\% accuracy of prediction, while holding bottle, pot, and chopsticks was still a little bit confusing to the model.

\subsubsection{Performance with different sizes of training data}
By manipulating the number of sessions used during the fine-tuning process, we generated the curve shown in Fig.~\ref{fig:cm} (b). In the UI setting, where no training data from the same participant was employed, the achieved accuracy reached 83.8\% (SD = 15.06\%). It is important to note that the participants exhibited variations in wrist thickness, hand size, and the preferred position for wearing the device. Specifically, some participants favored wearing the device in front of the ulnocarpal joint, while others preferred it directly on the ulnocarpal joint, and still others chose to wear it behind the ulnocarpal joint. Additionally, participants exhibited diverse approaches to interacting with objects. For example, due to the bottle's thickness, some participants with smaller hands could not grab its bottom portion and instead opted to grip the bottleneck, which was thinner and more accessible for them. Furthermore, the manner in which participants grasped the pan also displayed variations. Despite these individual discrepancies, \name{} consistently delivered promising results in recognizing and interpreting hand-object interactions.

Our observations from the fine-tuned results revealed an interesting trend. It became evident that even when the base model was trained using data from different individuals, incorporating a small amount of data from a new user resulted in significant performance improvements. For instance, when just 0.5 sessions (4 minutes) of data from the new user were added to the training set, the accuracy notably increased to 93.5\% (SD = 5.88\%) (Fig.~\ref{fig:cm} (b)). Furthermore, the variation among users' results decreased with fine-tuning. This serves as a strong validation of the fine-tuning process's effectiveness. It is reasonable to expect that with a more extensive dataset, we could achieve even better performance.

\subsubsection{User Experience Survey}
Based on the post-study survey results, the participants found \name{} comfortable to wear in a real-world setting (M = 6.0, SD = 0.77 on the Likert scale; 1 = extremely uncomfortable, 7 = extremely uncomfortable). Some participants (P05, P04, P06, P10) also commented that \textit{"it is just a general watch which I usually wear."}. Additionally, \name{} is lightweight (P03, P07, P08), and some participants reported that \textit{"I forgot that I was wearing the device during the experiment."} (P09, P11, P12). Overall, \name{} provided a good wearing experience. Furthermore, in our post-study survey, we asked participants whether they could hear any sound emanating from the device. All participants reported that they did not perceive any audible sound emitted from the device.

\section{Discussion}
% In this section, we discuss the potential opportunities of \name{}.

\begin{table*}[ht]
    \centering
    \caption{Comparison with other sensing methods.}
    \Description{Table.}
    \resizebox{2\columnwidth}{!}{
        \begin{tabular}{|c||c|c|c|c|c|c|c|c|c|}
        \hline
             & Technique & Form Factor & Hand Pose Tracking & Hand-Object Interaction Recognition & Thickness & Power & SI & UI \\
            \hline
            Digits \cite{kim2012digits} & IR Camera + IMU & Wristband & Continuous Hand Pose & \xmark & - & < 0.4 W & - & - \\
             &  &  & (Mean Errors All < 9\textdegree) &  &  &  &  & \\
            \hline
            DiscoBand \cite{devrio2022discoband} & Depth Sensor & Wristband & Continuous Hand Pose & Preliminary Exploration & < 1 cm & 3.6 W & MPJPE 17.87 mm & MPJPE 19.98 mm \\
             &  & (16 Depth Sensors) & (MJEPE 11.69 mm) &  &  &  &  & \\
            \hline
            FingerTrak \cite{hu2020fingertrak} & Thermal Camera & Wristband & Continuous Hand Pose & \xmark & 1.19 cm & 0.44 W & \xmark & \xmark \\
             &  & (4 Thermal Cameras) & (Average Angular Error 6.46\textdegree) &  &  &  &  & \\
            \hline
            Z-Ring \cite{waghmare2023zring, waghmare2023zpose} & Impedance & Ring + Armband & Continuous Hand Pose & 6 Objects & - & 2.4 W & - & \xmark \\
             &  & (1 VNA) & (Average Euclidean Error 7.2 mm) & (94.5\% Accuracy) &  &  &  & \\
            \hline
            EtherPose \cite{kim2022ether} & Impedance & Wristband & Continuous Hand Pose & \xmark & - & 4.5 W & \xmark & \xmark \\
             &  & (1 VNA) & (MPJPE 11.57 mm) &  &  &  &  & \\
            \hline
            Rudolph et al. \cite{rudolph2022sensing} & Capacitance & Wristband & \xmark & 6 Interactions & 0.7 cm & - & - & \xmark \\
             &  &  &  & (99\% Accuracy) &  &  &  & \\
            \hline
            AudioGest \cite{ruan2016audiogest} & Acoustic Signals & Commercial Laptop & Discrete Hand Gesture & \xmark & - & - & \checkmark & \checkmark \\
             &  &  or Smartphone & (6 Gestures, 96\% Accuracy) &  &  &  &  & \\
            \hline
            FingerIO \cite{nadakumar2016fingerio} & Acoustic Signals & Smartwatch & Continuous 2D Tracking & \xmark & - & last 4 hr & \checkmark & \checkmark \\
             &  & or Smartphone & (Accuracy 8 mm) &  &  &  &  & \\
            \hline
            BeamBand \cite{iravantchi2019beamband} & Acoustic Signals & Wristband & Discrete Hand Gesture & \xmark & 1 cm & 5 v, 400 mA & 89.4\% Accuracy & 51.7\% Accuracy \\
             &  &  & (6 Gestures, 94.6\% Accuracy) &  &  &  &  & \\
            \hline
            \name{} & Acoustic Signals & Continuous Hand Pose & Wristband & 12 Interactions & 0.6 cm & 0.0579 W & MJEDE 4.81 mm & MJEDE 12.2 mm \\
             &  &  & (MJEDE 4.81 mm) & (97.6\% Accuracy) &  &  & 97.6\% Accuracy & 83.8\% Accuracy \\
            \hline
        \end{tabular}
    }
    \label{tab:comparison}
\end{table*}

\subsection{Comparison with Other Sensing Methods on Hand Pose Recognition/Tracking}

Our study results showed that \name{} has a promising performance in continuously tracking hand poses and recognizing hand-object interactions. To help readers better situate the performance of \name{} with prior work, we highlight the key characteristics of prior work and \name{} in Table \ref{tab:comparison}. 

As the table shows, previous work on wristbands that were able to track hand pose continuously either require multiple form factor \cite{waghmare2023zring, waghmare2023zpose}, or consumes significant energy ranging from 0.4 W \cite{kim2012digits} to 4.5 W \cite{kim2022ether}, which is 10 to 100 times higher than \name{}. Furthermore, many of these works require training data from a new user \cite{hu2020fingertrak, waghmare2023zring, waghmare2023zpose, kim2022ether, rudolph2022sensing} or even a new session \cite{hu2020fingertrak, kim2022ether}. 

In comparison, \name{} can track hand pose continuously using a single wristband, consuming at most 1/10 of the prior works' energy (0.058 W) and providing promising tracking accuracy even without the training data from a new user. The closest prior work is DiscoBand \cite{devrio2022discoband}, which also tracks hand poses continuously and reports user-dependent and independent performance. In comparison, \name{} presents a better tracking performance, as shown in the table, with only 1/63 of their power signature. However, given the pose set is different, the comparison of performance may not be completely fair. therefore, we intend to present it as a reference for future directions.

\subsection{User Dependency of the Deep Learning model} \label{sec:user-dependency}

Data-driven wearable sensing system usually requires a significant amount of training data from a user before using the system. To improve the user experience, we strive to minimize user dependency to allow new users to access the system with ease. 

For hand pose tracking (Fig.~\ref{fig:pose-session}), when the new user does not provide any training data, \name{} still achieves 12.2mm MJEDE or 7.37\textdegree MJAE. With only 0.5 sessions of training data, \name{} obtains a significant performance improvement to 6.92mm MJEDE or 4.99\textdegree MJAE. The duration of 0.5 sessions is roughly one minute. In practice, one minute is close to the time needed to set up a fingerprint sensor or FaceID.

Hand-object interaction recognition achieved an average accuracy of 82.9\% for a new user without any training data. Similar to hand pose tracking, with only 0.5 sessions of training data, the accuracy significantly improved, reaching an average of 93.5\%. Although the data collection time is longer, taking four minutes, it remains comparable to the setup time of commonly used smart devices.

\name{} uses a pre-train-fine-tune scheme, which allows flexibility in incorporating new data and using a larger base dataset to improve performance. In the study, the base dataset was collected from 11 other users. In the future, the base dataset can be expanded at scale to improve \name{}'s performance further and eventually push \name{} towards a real user-independent system with strong performance.

\subsection{Hand Pose Tracking of Unseen Gestures}
\name{} includes two sets of hand gestures: 5 Simple Gestures and 10 Complex Gestures. Ideally, a hand-tracking technology would work on all hand poses. While it's difficult to evaluate all possible hand poses, testing the system's ability to estimate postures with unseen hand gestures (without training data) is a practical approach. It's worth noting that generalizing to unseen gestures is an extremely challenging task in the wearable community, and we have not seen any prior wearable hand-tracking system that conducted similar experiments. In this section, we present a preliminary analysis to provide insights into the promising potential of deploying \name{} in such real-world hand-tracking applications.

\begin{figure}[h]
  \centering
  \includegraphics[width=\columnwidth]{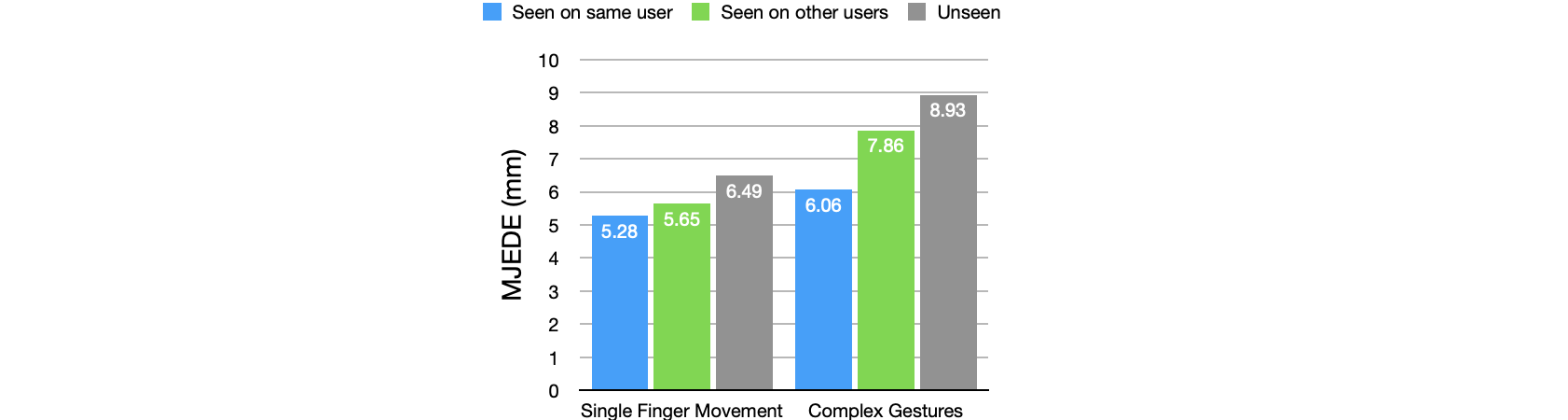}
  \caption{Performance of \name{} on unseen gestures. ``Seen on same user'': gestures in the testing set were also present in the training set from the same participant. ``Seen on other users'': gestures in the testing set were present in the training set but collected by different participants. ``Unseen'': gestures in the testing set were not present in the training set.}
  \Description{Bar chart with three series: seen on same user, seen on other users, unseen. V axis: MJEDE (mm). For single finger movement, values: 5.28, 5.65, 6.49. For complex gestures, values: 6.06, 7.86, 8.93.}
  \label{fig:unseen}
  \vspace *{-1.2em}
\end{figure}

In this experiment, we used the two gesture sets collected in the user study as the training and testing sets, respectively. 

We first established a baseline by evaluating the model's performance on data from the same gesture set for training and testing. In contrast to the previous experiment, we used a one-step training approach, where training data from all participants were combined to form the training set, and testing data from all participants were combined to form the testing set. In this experiment, the model had seen the same gesture (different instances) in both the training and testing datasets from the same participant. There was no overlap between the training and testing datasets. The results (Fig.~\ref{fig:unseen}), reveal that the MJEDE for the single finger movement gesture set and the complex gesture set was 5.28mm and 6.06mm, respectively (bar "Seen on Same User"). Please note that this performance is worse than that reported in Section~\ref{sec:gesture-performance} since the 2-step training was not applied.

We then evaluated the model performance on one gesture set that was trained on another different gesture set. In this case, gestures in the testing set were not present in the training set. The performance of the system decreased to 6.49mm and 8.93mm on single-finger movements (model trained using the data set on complex gestures) and complex gestures (model trained using the data set on single-finger gestures), respectively.

To improve the performance, we conducted experiments by including the same gesture collected from other users in the training set. In this way, the model still does not see any training data from the testing user on the target gestures. Results are shown in Figure~\ref{fig:unseen}, and the performance improved over the "Unseen" case, but there is still a gap between seen and unseen gestures. This indicates that incorporating other users' data on new gestures can improve performance.

The performance of \name{} decreased when estimating the poses of unseen hand gestures. However, to the best of our knowledge, this is the first experiment to predict unseen gestures in similar data-driven wearable hand posture tracking technologies. The performance was very encouraging (under 10mm) compared to other prior work. This again confirms the promising potential of this novel hand pose tracking technology for future deployment in real-world applications where hand poses vary significantly.

\subsection{Real-Time Inference}
We developed a real-time inference system on smartphones to further support various real-life applications. With the BLE module, \name{} enables real-time data transmission to a smartphone. We implemented the data processing and deep learning pipeline on the smartphone using PyTorch Mobile~\footnote{PyTorch Mobile~\url{https://pytorch.org/mobile/home/}}. The inference results can subsequently be transmitted to a laptop via WiFi for visualization or additional applications.

The delay of our real-time inference system spans from 0.3 seconds to 1 second for both hand pose tracking and hand-object interaction recognition. We logged the timestamp for each step during real-time inference testing, and ten random frames were chosen to be recorded. The average delay of each step is detailed in Table \ref{tab:delay}. The prediction delay of hand-object interaction recognition is larger since the window length and pixel of interest are larger, which results in a larger input size. The current bottleneck is in BLE transmission, stemming from a tradeoff between data throughput and latency. This challenge could be addressed by implementing data compression before transmission or relocating the computation process from the smartphone to the microcontroller. The fluctuating delay is primarily attributed to the performance of the smartphone and laptop. In instances of lower performance, e.g., when numerous other applications are running in the background of the smartphone or laptop, the prediction and rendering delay will noticeably increase. The delay can be mitigated by utilizing high-performance devices and further optimizing our algorithms on the operating system of the smartphone.

\begin{table*}[ht]
    \centering
    \caption{Delay breakdown in the two use cases.}
    \Description{Table. First row: Delay, BLE, Prediction, Echo Profile Calculation, WiFi, Rendering, Total. Second row: Hand Pose Tracking, 0.2567, 0.0578, 0.0018, 0.0550, 0.0716, 0.4429. Third row: Hand-Object Interaction Recognition, 0.2300, 0.2703, 0.0054, 0.0350, 0.0001, 0.5408}
    \begin{tabular}{|c||c|c|c|c|c|c|}
    \hline
        Delay (s) & BLE & Prediction & Echo Profile Calc & WiFi & Rendering & Total\\
        \hline
        Hand Pose Tracking & 0.2567 & 0.0578 & 0.0018 & 0.0550 & 0.0716 & 0.4429 \\
        \hline
        Hand-Object Interaction Recognition & 0.2300 & 0.2703 & 0.0054 & 0.0350 & 0.0001 & 0.5408 \\
        \hline
    \end{tabular}
    \label{tab:delay}
\end{table*}

\subsection{Integrating Hand Tracking and Hand-Object Interaction}
\name{} is able to track 3D hand pose as well as recognizing hand-object interactions. Both modules take in the exact same input and go through slightly different processes. While we did not evaluate how 3D hand tracking works when there are objects in hand, we demonstrate that it is possible to directly recognize the hand's interaction intentions with an end-to-end approach. In real use cases, it is possible to integrate the two modules together to allow for a more comprehensive understanding of the hand's activities. For instance, it is possible to use the hand-object interaction module to detect what object is interacting with the hand first. If no object is in the hand, the hand tracking module can be activated to understand the hand shape.

\subsection{Smartwatch Integration}
Our proposed technology is low-power and minimally-obtrusive, demonstrating the potential for integration into commodity smartwatches. However, several questions need to be addressed before it can be fully deployed on commodity devices.

\subsubsection{Form factor and Hardware}

\begin{figure*}[h]
  \centering
  \includegraphics[width=2\columnwidth]{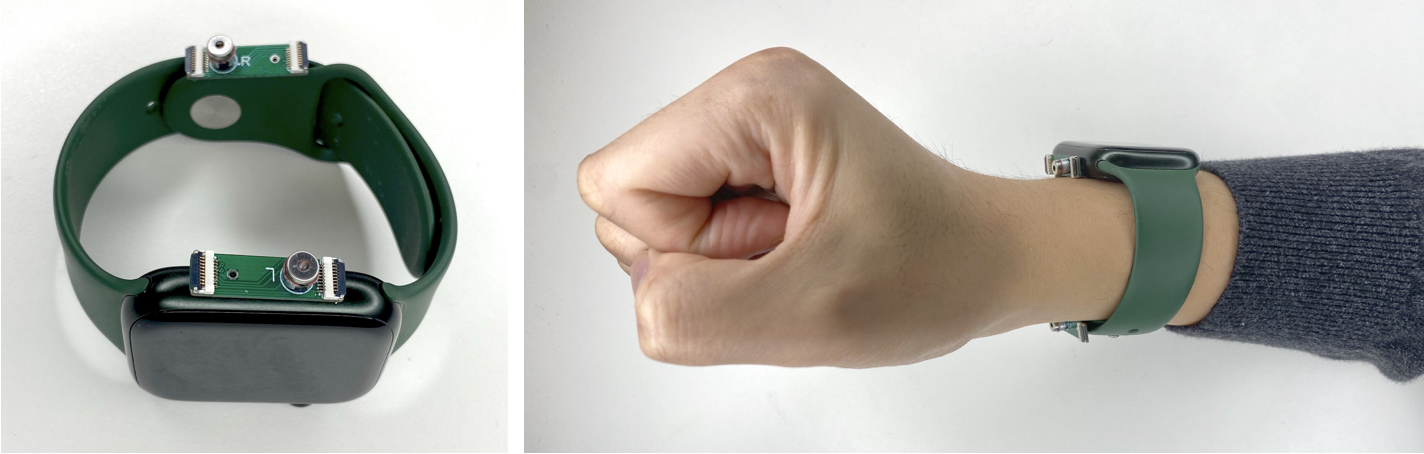}
  \caption{The mockup prototype integrating with an off-the-shelf smartwatch.}
  \Description{Tow photos of an Apple Watch combined with the sensing modules. On the left figure the Watch is on a desk. On the right figure the watch is worn on a wrist. The height of the sensing module is low that it blends with the thickness of the watch's main module and the strap.}
  \label{fig:mockup}
  \vspace *{-1.2em}
\end{figure*}

Unlike previous wrist-mounted sensing technologies that require cameras or sensors placed high above the skin, \name{} only requires two MEMS microphones and speakers placed 5mm above the skin. Therefore, integrating this sensing technology into future smartwatches is much easier. Figure~\ref{fig:mockup} illstrate how \name{} can be possibly integrated with Apple Watch Series 8 45mm~\footnote{\url{https://www.apple.com/apple-watch-series-8/}}. \name{} takes advantage of the thickness of the watch's body and the straps to keep the sensors minimally visible.

It's worth noting that many smartwatches on the market today, including the Apple Watch, already come equipped with speakers and microphones. This means that it's possible to adjust the position and orientation of these components in order to match our requirements. In our previous study, we found that this approach produced promising results.

Furthermore, the integration of the two pairs of microphones and speakers into the watch and the band, respectively, serves as a practical solution. Given the minimal cost, low energy consumption, and compact size of microphones and speakers, incorporating an extra pair into a smartwatch is well within the capabilities of watch manufacturers with the appropriate resources.  Moreover, integrating these sensing modules into existing hardware can lead to additional power savings, as our sensing solution can leverage the microprocessor and BLE module on the smartwatch. If we focus solely on the power consumption of the two pairs of microphones and speakers, the estimated power signature is as low as 10.0 mW.

\subsubsection{Privacy Protection}
\name{} uses 20-24 kHz acoustic sounds, which are generally inaudible to most people. Human conversations and most environmental sounds are usually distributed in low-frequency ranges that can be easily filtered out. However, in the current implementation of \name{}, all sounds under 25 KHz are recorded and transmitted to smartphones, which may expose potential privacy risks to users if the BLE transmission is hacked and raw audio is leaked, thereby damaging privacy. To mitigate this risk, several solutions can be adopted. Firstly, an analog band-pass filter can be implemented in the hardware to remove low-frequency sounds before they are converted into digital signals. Secondly, raw audio data transmission can be avoided in the processing system. This can be achieved by implementing a digital filter to remove audible frequencies in the data or by extracting features on the smartwatch and only transmitting processed features. Lastly, a complete local data processing pipeline can be employed on the user's personal devices, such as a smartphone, to avoid transmitting the data into the cloud. This approach has been demonstrated in a previous paper \cite{li2022eario}.

% \subsection{Wearing Experience}
% The design of \name{} incorporates a low-profile silicone wristband form factor to prioritize comfort. Silicone bands are a popular choice for wearable products, known for their robust durability, soft and pliable texture, and non-allergenic properties. In the construction of the 3D-printed sensor support case, we opted for TPU due to its flexibility. According to feedback from the participants in the user study, this material provided a comfortable experience similar to that of everyday watches. Additionally, the lightweight and comfortable design made participants almost forget they were wearing the device. Furthermore, the low profile of \name{} holds the potential for greater social acceptance.

\section{Limitations and Future Work}
\subsection{Cloth Coverage} 
One limitation shared by \name{} and many other wrist-mounted sensing methods based on cameras \cite{hu2020fingertrak} is the potential degradation of system performance when the sensing unit on the wristband is obscured by clothing. In such cases, the acoustic signals may be obstructed by the fabric and fail to reach the hands, particularly when users are wearing long-sleeved clothing. This limitation imposes a requirement for users of our system to wear clothing that does not cover the wrist. In some cases, clothing might obstruct some sensing units partially while leaving others unaffected. In future research, this feature could be leveraged to enhance the algorithm's robustness or integrated with other sensing methods to mitigate occlusion issues arising from clothing.

\subsection{Performance under Intense Movement and Holding Objects} 
While \name{} successfully tracks both hand postures and hand-object interactions, its performance in tracking hand postures while holding objects or during intense movements remains unexplored. This is a challenging issue encountered by many hand activity tracking systems. We believe that performing intense movement and holding objects could introduce distinctive acoustic echo profiles. To address this, further investigation is needed, potentially involving the collection of additional training data specific to these scenarios. Future research may delve into this aspect to enhance the system's capabilities.

\subsection{Hand-Object Interaction Contexts}
In study 2 (Section~\ref{sec:evaluation-2}, we evaluated \name{}'s performance in recognizing hand-object interactions within a kitchen setting. However, this does not mean that \name{} can only be used in the kitchen. Pilot studies involving one researcher testing the system in different contexts were conducted. The interactions that have been tested included writing, typing, scrolling, cutting, and more. Most of the tested interactions demonstrated impressive results, achieving accuracy rates exceeding 90\% when training and testing with the data from the same user. However, it is challenging for \name{} to distinguish holding objects sharing similar shapes and grabbing poses, e.g., fork and spoon, pencil and marker, and hot glue gun and drill. In addition, objects that are fully in hand can not be recognized as well. These include AirPods, erasers, and candies. Future research could encompass more expansive contexts to further gauge \name{}'s adaptability and robustness across a broader spectrum of everyday scenarios. Also, multimodel approaches could be deployed to extend the capability further. 
\section{Conclusion}
We present the design, implementation and evaluation of \name{}, the first wristband that can both track 3D hand poses continuously and recognize hand-object interactions. Two user studies with 24 participants in total demonstrate \name{}'s capability and robustness in these two tasks. \name{} operates at 56.9mW while maintain a low-profile minimally-obtrusive form factor. With further optimization, we believe that it is promising to deploy \name{} at scale.

\begin{acks}

\end{acks}

\bibliographystyle{ACM-Reference-Format}
\bibliography{100_References}

% \appendix









\end{document}